\documentclass[10pt,journal,twoside,final]{IEEEtran}
\usepackage{diagbox}
\usepackage{graphicx}
\graphicspath{{./Pictures/}} 
\usepackage{tablefootnote}
\usepackage{amsmath}
\usepackage{threeparttable}
\usepackage{url}
\usepackage{array}
\usepackage[linesnumbered,ruled,vlined]{algorithm2e}
\SetKwInput{KwInput}{Input}
\SetKwInput{KwOutput}{Output} 
\newcolumntype{L}[1]{>{\raggedright\let\newline\\\arraybackslash\hspace{0pt}}m{#1}}
\newcolumntype{C}[1]{>{\centering\let\newline\\\arraybackslash\hspace{0pt}}m{#1}}
\newcolumntype{R}[1]{>{\raggedleft\let\newline\\\arraybackslash\hspace{0pt}}m{#1}}
\usepackage{amssymb}

\DeclareMathOperator*{\argmin}{arg\,min}
 \usepackage{booktabs}

\usepackage{multirow}
\usepackage{array}

\usepackage{bbding}
\usepackage{bigstrut}
\usepackage{cite}

\ifCLASSOPTIONcompsoc
\usepackage[caption=false,font=normalsize,labelfont=sf,textfont=sf,subrefformat=parens,labelformat=parens]{subfig}
\else
\usepackage[caption=false,font=footnotesize,subrefformat=parens,labelformat=parens]{subfig}
\fi

\usepackage{color,xcolor,colortbl}
\newcommand{\etal}{\textit{et al. }}
\newcommand{\mj}[1]{{\color{black}#1}}

\usepackage{algpseudocode}
\algnewcommand\algorithmicinput{\textbf{INPUT:}}
\algnewcommand\INPUT{\item[\algorithmicinput]}
\algnewcommand{\algorithmicoutput}{\textbf{OUTPUT:}}
\algnewcommand\OUTPUT{\item[\algorithmicoutput]}
\algrenewcommand{\algorithmiccomment}[1]{\hskip3em$\%$ #1}

\begin{document}

\title{Adversarial Attacks Against Deep Learning-Based Radio Frequency Fingerprint Identification}

\author{Jie~Ma,
	Junqing~Zhang,~\IEEEmembership{Senior Member,~IEEE,}
    Guanxiong~Shen,
	Alan~Marshall,~\IEEEmembership{Senior Member,~IEEE},
	and Chip-Hong~Chang,~\IEEEmembership{Fellow,~IEEE}

\thanks{
Manuscript received xxx; revised xxx; accepted xxx. Date of publication xxx; date of current version xxx. 
The work of J. Zhang was in part supported by the UK Engineering and Physical Sciences Research Council (EPSRC) under grant ID EP/V027697/1 and EP/Y037197/1 as well as in part by the UK Royal Society Research Grants RGS$\backslash$R1$\backslash$231435.
The work of J.~Ma was supported by the China Scholarship Council (CSC), under Grant 202006470007. The work of C. H. Chang was supported by the National Research Foundation, Prime Minister’s Office, Singapore, under its Campus for Research Excellence and Technological Enterprise (CREATE) IN-CYPHER programme.
This paper was presented in part at the IEEE ICC 2023. For the purpose of open access, the authors have applied a Creative Commons Attribution (CC BY) licence to any Accepted Manuscript version arising. 
The review of this paper was coordinated by xxx. 
\textit{(Corresponding author: Junqing Zhang.)}}

\thanks{J. Ma, J.~Zhang, and A.~Marshall are with the School of Computer Science and Informatics, University of Liverpool, Liverpool, L69 3GJ, United Kingdom. (email: Jie.Ma@liverpool.ac.uk; Junqing.Zhang@liverpool.ac.uk; Alan.Marshall@liverpool.ac.uk)}
\thanks{G.~Shen is with the School of Cyber Science and Engineering, Southeast University, Nanjing, China. (email: gxshen@seu.edu.cn)}
\thanks{C.-H.~Chang is with the School of Electrical and Electronic Engineering, Nanyang Technological University, Singapore. (email: echchang@ntu.edu.sg)}
\thanks{Color versions of one or more of the figures in this paper are available online at http://ieeexplore.ieee.org.}
\thanks{Digital Object Identifier xxx}	
}

\maketitle

\begin{abstract}
Radio frequency fingerprint identification (RFFI) is an emerging technique for the lightweight authentication of wireless Internet of things (IoT) devices. RFFI exploits deep learning models to extract hardware impairments to uniquely identify wireless devices. Recent studies show deep learning-based RFFI is vulnerable to adversarial attacks. However, effective adversarial attacks against different types of RFFI classifiers have not yet been explored. In this paper, we carried out a comprehensive investigations into different adversarial attack methods on RFFI systems using various deep learning models. Three specific algorithms, fast gradient sign method (FGSM), projected gradient descent (PGD), and universal adversarial perturbation (UAP), were analyzed. The attacks were launched to LoRa-RFFI and the experimental results showed the generated perturbations were effective against convolutional neural networks (CNNs), long short-term memory (LSTM) networks, and gated recurrent units (GRU). We further used UAP to launch practical attacks. Special factors were considered for the wireless context, including implementing real-time attacks, the effectiveness of the attacks over a period of time, etc. Our experimental evaluation demonstrated that UAP can successfully launch adversarial attacks against the RFFI, achieving a success rate of 81.7\% when the adversary almost has no prior knowledge of the victim RFFI systems.
\end{abstract}
	

\section{Introduction}
\IEEEPARstart{r}{adio} frequency fingerprint identification (RFFI) is a device authentication technique that is suitable for Internet of Things (IoT) devices~\cite{zhang2021radio}.
The transmitter chain of a wireless device consists of numerous analog components such as mixers, oscillators, power amplifiers, antennas, etc. 
These components deviate slightly from their nominal specifications due to the inevitable manufacturing process variations, i.e., hardware imperfections such as I/Q imbalance, carrier frequency offset (CFO), and power amplifier non-linearity~\cite{li2022radio,hanna2020open, sankhe2019no}. 
Although these hardware impairments will not affect communication transmission, they will slightly impact the waveform of wireless transmission~\cite{zhang2021radio}. 
An RFFI system is designed to identify wireless devices by analyzing the unique distortion of the received waveform caused by the imperfections of hardware components.

\mj{Traditionally, RFFI approaches manually extract specific features, such as CFO~\cite{cfo1, cfo2} or I/Q imbalance~\cite{brik2008wireless}, for classification. However, such handcrafted methods may overlook the complex and intertwined nature of these imperfections.
Recent research has increasingly adopted deep learning (DL) techniques to learn device-specific features from raw waveforms. 
DL-based RFFI systems leverage the powerful feature extraction and classification capabilities of models, such as convolutional neural networks (CNNs)~\cite{jian2020deep,xie2021generalizable,qian2021specific,rajendran2022rf,wang2022radio,jian2021radio,jagannath2023embedding,shen2023towards,kong2025deepcrf}, long short-term memory (LSTM)~\cite{shen2021jsac, he2020cooperative} and gated recurrent units (GRUs)~\cite{roy2019rf}. They effectively capture the unique characteristics embedded in wireless signals, enabling device identification without relying on manually selected hardware imperfections.}

As the core of a DL-based RFFI system, the security of the DL model should be ensured. However, the latest research shows that DL is vulnerable to adversarial machine learning (AML)~\cite{adesina2022adversarial}. 
AML can occur at both training and inference stages. For example, the backdoor attacks can add malicious triggers in the training stage~\cite{zhao2025tmc}. This paper focuses on the attacks at the inference stage, as they can be easily launched due to the broadcast nature of wireless transmissions. Specifically, the adversarial attack, also known as the evasion attack, can inject subtle perturbations to the inputs in the inference stage, misleading the DL model to make an incorrect prediction.
While AML has been widely studied in computer vision~\cite{Dong_2018_CVPR, moosavi2016deepfool, Moosavi-Dezfooli_2017_CVPR, Zhang_2021_ICCV}, it has recently been applied to wireless domains~\cite{adesina2022adversarial}. 
Research on adversarial attacks in wireless communications includes localization~\cite{xiao2023over,liu2023exploring}, channel prediction~\cite{liu2023exploring, huang2021wars, zhou2022wiadv, yang2022securesense},  modulation recognition~\cite{sadeghi2019physical, sadeghi2018adversarial, sagduyu2019adversarial, flowers2019evaluating, kim2021channel,lin2020threats,lin2020adversarial} and WiFi sensing~\cite{yin2025evasion}. 
Applying AML to wireless signals requires additional considerations due to the unique characteristics of wireless communication, e.g., channel fading~\cite{kim2021channel, hameed2021} and clock offsets~\cite{liu2023exploring}, etc. 

Specifically, to adversarial attacks against DL-based RFFI, there are also some initial works.
Bao~\etal~\cite{bao2021threat} examined the effects of adversarial attacks on CNNs, but not on other DL models, and the signals were collected from a wired setup.
Restuccia~\etal~\cite{ restuccia2020generalized} investigated adversarial attacks on I/Q samples, and both Wi-Fi and ADS-B datasets were studied. However, they focused on white-box attacks, which assume the adversary has all the necessary knowledge, and it is less practical in realistic scenes.
\mj{Liu~\etal~\cite{liu2023robust} proposed an attack method that takes the practical fading channels into consideration, but the results were derived from simulation, and the classification accuracy of the victim system is insufficient, e.g., the accuracy without the attack is only 61\%. 
Papangelo~\etal~\cite{papangelo2024adversarial} studied adversarial attack and countermeasures for image-based RFFI. 
Li~\etal~\cite{li2024slpa} designed a single-line pixel attack by changing a line of the image-based spectrogram. However, the adversary needs to query the victim system, which may not be practical because the victim will not be cooperative. 
In~\cite{yalin2025iotj}, the authors investigate adversarial attacks on multi-task LoRa-based RFFI systems that simultaneously classify legitimate devices and detect rogue devices, conducting experiments with two legitimate transmitters and two rogue devices.}
In brief, the adversarial attack against RFFI is still nascent, and many practical challenges remain unanswered and a systematic investigation is missing.

This paper carried out a comprehensive investigation of adversarial attacks against DL-based RFFI systems. 
Three attack algorithms are investigated, namely fast gradient sign method (FGSM)~\cite{goodfellow2014explaining}, projected gradient descent (PGD)~\cite{madry2018towards}, and universal adversarial perturbation (UAP)~\cite{Moosavi-Dezfooli_2017_CVPR}. This paper also examined adversarial attacks in practical scenarios by carefully involving unique features in the wireless context. 
The victim system used in this study is a LoRa-RFFI testbed consisting of a USRP N210 software-defined radio (SDR) receiver and ten commercial-off-the-shelf (COTS) LoRa transmitters. Three types of classification neural networks (NNs) are investigated, namely CNN, LSTM, and GRU.
Our main contributions are detailed as follows:
\begin{itemize}
\item FGSM and PGD are used to generate a unique perturbation for each sample and can severely degrade the classification performance. In a CNN-based RFFI system, 87.2\% and 96.2\% of packets were shown to be successfully misclassified by FGSM and PGD attacks, respectively.  
A targeted attack can be launched by PGD via superimposing elaborately generated perturbations to the inputs and misleading the DL model to output a
specific label set by the adversary. Our experiments can
make up to 98.9\% of packets to be classified from the
target device.
\item We employed UAP to generate a universal perturbation for all input samples.
UAP can maintain effectiveness even when the victim's dataset and DL model are unknown. 
In such a case, 88.2\% of packets are misclassified when the victim is an LSTM-based RFFI system.

\item We further revealed practical adversarial attacks using UAP under wireless contexts, considering different surrogate DL models, attack time, surrogate devices, and real-time attacks.
It is verified that UAP preserves effectiveness even if all the aspects mentioned are combined, e.g., 81.7\% packets are misclassified when the victim employs a CNN-based RFFI system.
\end{itemize}
In our previous work, we presented non-targeted and targeted attacks in the white-box setting~\cite{majie}. In this paper, we considerably extend and complete this work by studying adversarial attacks in grey/black-box settings and more practical scenarios related to wireless communication.

The rest of the paper is organized as follows. Section~\ref{sec:RFFI_PRIMER} introduces the system of RFFI. Section~\ref{sec:sys_overview} introduces the adversarial attack against RFFI and Section~\ref{sec:attackmethod} presents perturbation generation methods.
Section~\ref{sec:lora-rffi} introduces the design details of the RFFI system and Section~\ref{sec:experi_eva} evaluates the effectiveness of adversarial attacks on different DL models in the RFFI system.
Section~\ref{sec:practical_tests} considers more factors that could impact the effectiveness of adversarial attacks in practical wireless scenarios. Section~\ref{sec:conclusion} concludes the paper.

\section{RFFI Overview}\label{sec:RFFI_PRIMER}
\begin{figure}[!t]
    \centering
    \includegraphics[width=3.4in]{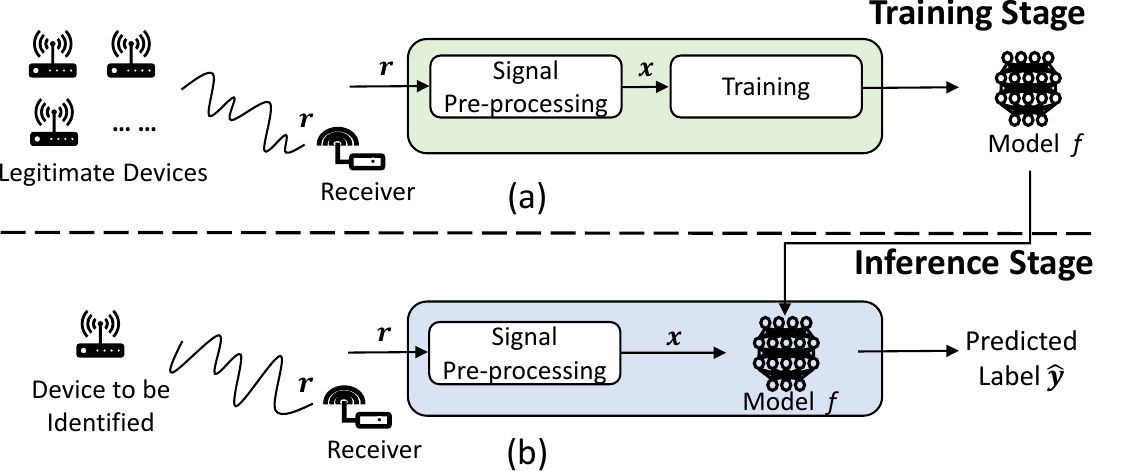}
    \caption{Overview of an RFFI system. (a) Training stage. (b) Inference stage.}
    \label{fig:RFFIsys}
\end{figure}

A typical DL-based RFFI system is illustrated in Fig.~\ref{fig:RFFIsys}, which consists of training and inference stages. There are $N$ devices under test (DUTs) and a receiver for capturing their wireless waveforms for identification.

\subsection{Training Stage}
A receiver will collect numerous wireless signals $r$ from $N$ legitimate devices, as shown in Fig.~\ref{fig:RFFIsys}(a). The collected signals will be pre-processed and converted into a channel-independent data representation before they are fed into the DL-model $f$ for training. The pre-processing operations include synchronization, normalization, CFO compensation, preamble extraction, etc, which are elaborated in our prior work~\cite{shen2021jsac}. The processed signals and their corresponding transmitter identities (e.g., device indices) constitute the training dataset $\mathcal{X}^{train}$,  given as
\begin{equation}
    \mathcal{X}^{train} = \{(x_i,y_i)\}_{i=1}^{I}, 
\end{equation}
where $x_i$ is the $i^{th}$ training instance, $y_i$ is its label and $I$ is the number of instances in the training dataset.

The learnable parameter set $\theta$ of an NN classifier is updated by iterating over the set of training data through the network to minimize the loss between the network outputs and the correct labels. The process can be mathematically expressed as
\begin{equation}\label{eqn:train}
    \theta=\mathop{\argmin}_{\theta} \sum_{(x, y)\in \mathcal{X}^{train}} \mathcal{L}(f(x;\theta), y),
\end{equation}
where $\mathcal{L}(\cdot)$ is the cross-entropy loss function and $f(\cdot;\theta)$ denotes the NN model.

    

\subsection{Inference Stage}
The inference stage is illustrated in Fig.~\ref{fig:RFFIsys}(b). 
The received signal $r$ undergoes the same pre-processing as the training data before it is fed into the pre-trained DL model $f(\cdot;\theta)$ for classification. The classifier outputs a prediction  
\begin{equation}
    \hat{y} = f(x;\theta),
\end{equation}
where $\hat{y}$ is a label predicted by $f$ for the most likely transmitter of the input signal $r$.



\section{Adversarial Attacks on RFFI Systems}\label{sec:sys_overview}
This section introduces the background of adversarial attacks and an overview of the attacks on RFFI systems, including attack design, categorization of attack methods, and evaluation metrics. For ease of exposition, we refer to an RFFI system under attack as a victim system.

\mj{\subsection{Adversarial Attacks}
\begin{figure}[!t]
    \centering
    \includegraphics[width =3.4in]{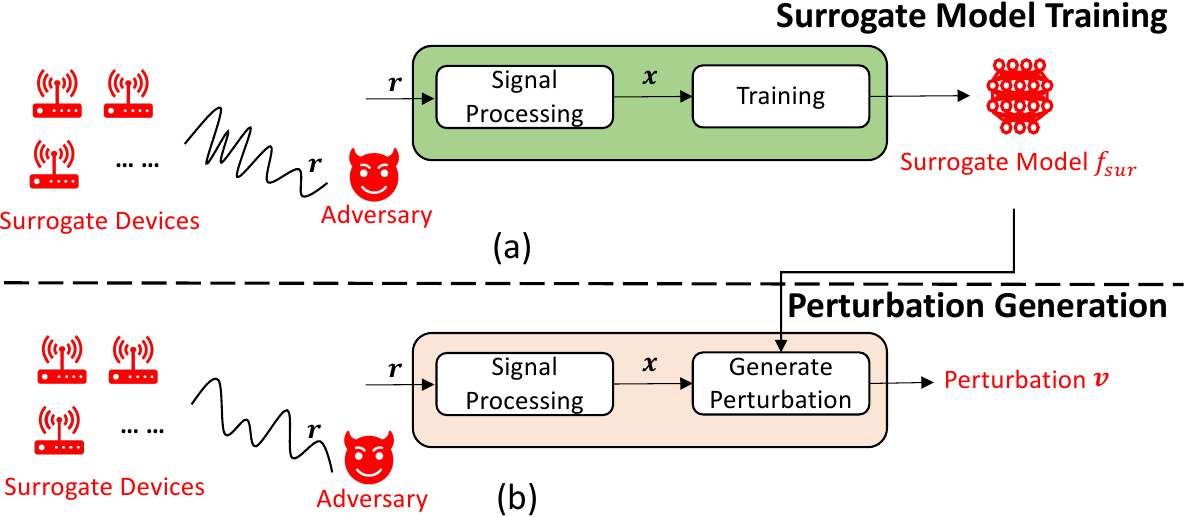}
    \caption{The procedure of generating adversarial perturbation. (a) The adversary trains a surrogate model. (b) The adversary uses the trained surrogate model to generate the perturbation.}
    \label{fig:RFFI_with_attack}
\end{figure}

Adversarial attacks aim to cause misclassification in a DL classifier by adding carefully crafted perturbations to its input. 
Directly generating such perturbations against a victim system is challenging, as the adversary often lacks access to the victim’s data and DL model. 
To overcome this, the surrogate devices and a surrogate model are employed by the adversary.

\subsubsection{Surrogate Devices and Surrogate Model}
Surrogate devices are employed since it will be challenging for the adversary to collect sufficient signals from the victim devices. These devices operate under the same RF protocol as the victim devices, thereby providing datasets for model training and perturbation generation.

Similarly, surrogate models are adopted because the adversary lacks access to the victim’s DL model. 
A surrogate model serves as a proxy for the victim model,
enabling the generation of perturbations without access to the victim.

\subsubsection{Surrogate Model Training}
As shown in Fig.~\ref{fig:RFFI_with_attack}(a), the adversary first collects RF signals from surrogate devices to train a surrogate model $f_\emph{sur}$. 
This step is similar to the training process of the victim system in Fig.~\ref{fig:RFFIsys}(a). 
However, depending on the knowledge of the adversary, the training dataset and the DL model used for $f_\emph{sur}$ may be different from the victim system, which will be elaborated in Section~\ref{sec:category}.

\subsubsection{Perturbation Generation}
As depicted in Fig.~\ref{fig:RFFI_with_attack}(b), the perturbation $v$ is generated based on the surrogate model $f_{sur}$ and the signals collected from the surrogate devices. 
This paper investigates three perturbation
generation algorithms, namely FGSM, PGD, and UAP, which will be explained in Section~\ref{sec:attackmethod}.
\subsubsection{Adversarial Attacks Implementation}
Once the perturbation is generated, the adversary can use it to launch an adversarial attack. 
To directly analyze the impact of such attacks on the classifier’s predictions and to represent the worst-case scenario, we focus on digital attacks~\cite{flowers2019evaluating, sadeghi2018adversarial, lin2020threats,lin2020adversarial} in this paper, where the perturbation is superimposed on the classifier (DL model) directly.
In this setting, the adversarial attack can be efficiently carried out by the transformed $x^{\prime}$, as shown in Fig.~\ref{fig:attack_against_RFFI}.
The transformed signal $x^{\prime}$ is given by
\begin{equation}\label{eqn:nosnr}
x^\prime = x + v,
\end{equation}
where $x$ and $v$ denote the pre-processed signals of the received signal $r$ and the perturbation $r_v$, respectively.
The output $\hat{y}^\prime$ becomes
\begin{equation}
    \hat{y}^\prime = f(x^\prime;\theta).
\end{equation}
\begin{figure}[!t]
    \centering
    \includegraphics[width =3.3in]{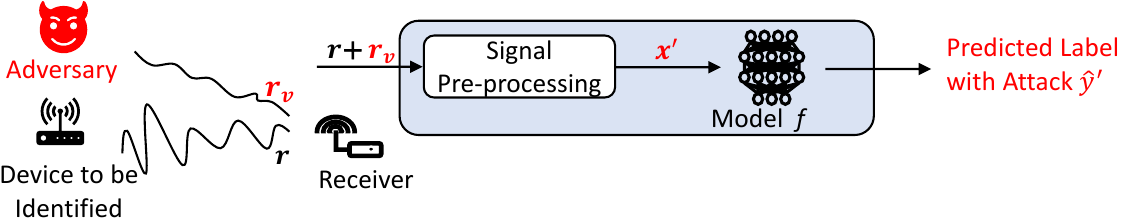}
    \caption{Adversarial example inputs to the DL classifier at the inference stage of an RFFI system.}
    \label{fig:attack_against_RFFI}
\end{figure}
}

\mj{\subsection{Threat Model}\label{sec:category}
The threat model is characterized by three key aspects: the attack stage, the
knowledge, and the goal of the adversary. 
\subsubsection{Attack Stage}
It specifies the phase during which the adversary can take action. 
Adversarial attacks typically occur at the inference stage, therefore, the perturbations are also introduced during inference.

\subsubsection{Knowledge}\label{sec:attacks}
According to the knowledge of the adversary, the adversarial attacks can be categorized into white-box, grey-box, and black-box attacks~\cite{li2020adversarial}.

\begin{itemize}

    \item \textbf{White-box attack (WB):} the adversary has the full knowledge of the victim system, including the dataset and DL model.
    \item \textbf{Grey-box attack (GB):} the adversary has partial knowledge of the victim. 
    In a ``GB1" setting, the model's information is revealed, but the dataset is unavailable. The known DL model and surrogate dataset can be used to generate the perturbations.
    In a ``GB2" setting, the information about the dataset is disclosed, but the structure of the DL model $f$ is unknown. A surrogate model $f_\emph{sur}$ will be used in place of $f$ for the generation of the perturbations.

    \item \textbf{Black-box attack (BB):} the adversary has no knowledge about the victim system. The perturbations will be generated using the surrogate dataset and surrogate DL model.

\end{itemize}

\subsubsection{Goal} According to the adversary’s goal, adversarial attacks can be categorized into non-targeted and targeted attacks. 
In the non-targeted attack, the aim is to make the prediction of an adversarial example different from the original label, i.e.,
    \begin{equation}
    \hat{y}' \ne y.
    \end{equation}
    
The goal of the targeted attack is to have all adversarial examples predicted to be the target label, i.e.,
    \begin{equation}\label{eqn:TAadv}
    \begin{aligned}
      \hat{y}' = y_{tar},
    \end{aligned}
    \end{equation}
where $y_{tar}$ is the target label set by the adversary.}

\subsection{Metrics}
The perturbation-to-signal ratio (PSR) and the success rate (SR) are two critical metrics of adversarial attacks.

PSR is the ratio of the power of the perturbation, $P_v$, to the power of the original signal, $P_x$, expressed as
\begin{equation}\label{eqn:psr}
    PSR=10 \log_{10} (\frac{P_{v}}{P_{x}}).
\end{equation}

SR is defined as the probability of success of an adversarial attack. The SR of a non-targeted attack is calculated by
\begin{equation}\label{eqn:foolingrate}
     SR = \frac{\sum_{m=1}^{M} \zeta \left ( \hat{y}'_m \ne y_m \right ) }{M},
\end{equation}
where $M$ is the total number of test signals, $\hat{y}'_m$ is the predicted label of the $m^{th}$ adversarial example and $y_m$ is the ground truth label of its clean sample. ${\zeta}(\cdot)$ is an indicator function that returns one if its argument is true and zero otherwise.

\mj{In the case of a targeted attack with the target being $y_{tar}$, its SR is calculated by
\begin{equation}
SR = \frac{1}{M}\sum_{m=1}^M \zeta\!\left(\hat{y}_m^{\prime}, y_{\mathrm{tar}}\right);
\end{equation}
where
\[
\zeta\!\left(\hat{y}_m^{\prime}, y_{\mathrm{tar}}\right) =
\begin{cases}
1 & \text{if } \hat{y}_m^{\prime} \text{ equals to } y_{\mathrm{tar}}\\
0 & \text{otherwise.}
\end{cases}
\]
}


\section{Perturbation Generation Methods}\label{sec:attackmethod}

Among the white-box adversarial examples generation methods, FGSM and its iterative version, PGD are the most popular methods. They are also the basis for many improved evasion attacks. These attacks are useful for assessing a DL model's resilience against adversarial attacks under the worst-case, white-box scenario, where all the knowledge about the victim model can be used for the perturbation generation. In contrast, UAP is applicable to the grey/black-box setting as it is input-agnostic and it makes no assumption on knowledge about the victim model. As UAP uses the geometric characteristics of the classifier to calculate the minimum perturbation required to cross the nearest decision boundary, the adversary may have to train its own surrogate model. 


\subsection{Fast Gradient Sign Method (FGSM)}
FGSM~\cite{goodfellow2014explaining} is a non-iterative approach to produce perturbations that can reliably cause a wide variety of NNs to malfunction with a one-step projected gradient. 
FGSM calculates the gradient direction of the loss function relative to the NN input to determine the direction of the generated perturbation with minimum cost. Only the sign of the gradient is used by FGSM. 

Specifically, an adversarial example for a non-targeted attack is generated by
\begin{equation}\label{eqn:fgsmnon}
   x^{\prime}=x+\varepsilon \cdot \operatorname{sign}\left(\nabla_{x} \mathcal{J}(f(x;\theta), y)\right),
\end{equation}
where $\varepsilon$ denotes a parameter that controls the amplitude of the perturbation, $\nabla_x$ indicates the gradient of the model for a clean sample $x$, $\mathcal{J}(\cdot)$ denotes the loss function and $\operatorname{sign}(\cdot)$ indicates the gradient direction. $\operatorname{sign}(\cdot)$ returns 1 ($-1$) if the value of the gradient direction is greater (smaller) than 0.

For targeted attacks, it is generated by
\begin{equation}\label{eqn:fgsmtar}
    x^{\prime}=x-\varepsilon \cdot \operatorname{sign}\left(\nabla_{x} \mathcal{J}(f(x;\theta), y_{tar})\right).
\end{equation}

\subsection{Projected Gradient Descent (PGD)}
PGD generates adversarial examples by an iterative method with random initialization. It overcomes the underfitting problem encountered by FGSM due to the linear approximation of the decision boundary around the data point. Unlike FGSM, which applies a one-time perturbation of input data in the direction of the gradient, PGD iteratively applies FGSM and projects the perturbed sample back to the norm multiple times with the step size scaled by the total perturbation bound. Hence, it has a much larger space for perturbation exploration than FGSM to produce more effective adversarial examples with less conspicuous perturbations.

\mj{The procedure of PGD is given in Algorithm~\ref{algorithm:pgd}, where $\left \| \cdot \right \| _p$ is the norm of the perturbation (such as $L_2$ or $L_\infty$). $\operatorname{Clip}_{x, \alpha}(\cdot)$ is a clipping function that limits the argument to the range $[x-\alpha, x+\alpha]$.}

\begin{algorithm}[!t]
    \DontPrintSemicolon
      \mj{\KwInput{signal ${x}$, model $f(\cdot;\theta)$, label $y$ or $y_{tar}$, loss function ${\mathcal{J}(\cdot)}$, size of the perturbation $\varepsilon$, the size on each iteration step $\alpha$, number of iteration $K$ }
      \KwOutput{an adversarial example $ x_{K}^{\prime}$ with $\left\|x_{K}^{\prime}-x\right\|_{\text {p}} \leq \varepsilon$}
    Initialize $x_0^{\prime} \leftarrow x$ 
    
    \For{$k$ in {1, 2, …, $K$}}{
    \If{Non-targeted attack}{
    $x_{k}^{\prime}\leftarrow\operatorname{Clip}_{x, \alpha}\left\{x_{k-1}^{\prime}+\alpha \cdot \operatorname{sign}\left(\nabla_{x} \mathcal{J}(f(x;\theta), y)\right)\right\}$
    }

      \ElseIf{Targeted attack}{$x_{k}^{\prime}\leftarrow\operatorname{Clip}_{x, \alpha}\left\{x_{k-1}^{\prime}-\alpha \cdot \operatorname{sign}\left(\nabla_{x} \mathcal{J}(f(x;\theta), y_{tar})\right)\right\}$
      }
    }
    
       \Return $ {x_K^{\prime}}$
    \caption{PGD Algorithm}\label{algorithm:pgd}
    \label{alg:pgd}}
\end{algorithm}

\subsection{Universal Adversarial Perturbation (UAP)}
FGSM and PGD require full knowledge of the victim model parameters and architecture to generate a unique perturbation for each attack sample~\cite{goodfellow2014explaining, madry2018towards, Dong_2018_CVPR}. Such a white-box setting may not be always feasible. UAP~\cite{Moosavi-Dezfooli_2017_CVPR} is able to generate adversarial examples when knowledge about the model parameters and architecture is not available~\cite{sadeghi2019physical,sadeghi2018adversarial,sagduyu2019adversarial}. More importantly, the perturbation generated by UAP is input independent, which means that the same perturbation needs only to be generated once and is applicable to all the samples. This is in stark contrast to FGSM and PGD, where a perturbation has to be generated individually for each sample.

\begin{algorithm}[!t]
    \DontPrintSemicolon
    \KwInput{signals $x$, model $f_{sur}(\cdot;\theta)$, size of the perturbation $\varepsilon$, desired SR $\delta$, number of iteration $T$}
    \KwOutput{perturbation $v$}
    Initialize $v \leftarrow 0 , t \leftarrow 0 $
    
    \While{$SR \le \delta $ and $t \le T$} {

                   {\For{each sample $x_m\in x$}
                            { \While{$ f_{sur}(x_m+v;\theta) \leftarrow f_{sur}(x_m;\theta) $}{
                        $v \leftarrow v + \mathcal{G}(x_m+v,f_{sur}(\cdot;\theta))$,
                        \\\text {subject to }$\left\|v\right\|_{\text {p}} \leq \varepsilon$
              	        	}
              	        	}
                        Inference: using (4) and (6) to get $y_m$ and $\hat{y}'_m$, respectively

                        $t = t + 1$
                        
                        $SR = \frac{\sum_{m=1}^{M} \zeta \left ( \hat{y}'_m \ne y_m \right ) }{M}$

                    }
    }
    \Return $v$
    \caption{UAP Algorithm}\label{algorithm:uap}
\end{algorithm}

The UAP algorithm is shown in Algorithm~\ref{algorithm:uap}. \mj{In each iteration, the current perturbed sample $(x_m+v)$ is sent to the decision boundary by a UAP generation method $\mathcal{G}(\cdot,f(\cdot;\theta))$, such as DeepFool~\cite{moosavi2016deepfool}, Cosine~\cite{Zhang_2021_ICCV}, etc. 
Here, we use DeepFool~\cite{moosavi2016deepfool} because it provides an effective search for the decision boundary, and is deterministic, yielding identical perturbations under the same conditions.}
The perturbation $v$ is computed in Step 5, and its norm is checked against the distortion bound in Step 6. This is to regulate the extent of perturbation. The perturbation $v$ for the current sample will be updated until the current perturbed sample achieves the desired misclassification on the surrogate DL model or reached the loop limit $T$. This process is repeated for the next sample until the $SR$ computed in Step 8 exceeds the preset threshold $\delta$. At which point, the program terminates and $v$ is produced as a UAP.

\mj{\section{DL-based RFFI System for LoRa}\label{sec:lora-rffi}
Although adversarial attacks can be applied to any DL-based RFFI system, this section focuses on the design and implementation of a LoRa signal-driven RFFI system, which is employed as the testbed for evaluating adversarial attacks in the subsequent sections.}

\subsection{LoRa Primer}
\begin{figure}[!t]
\centering
\subfloat[]{\includegraphics[width=1.54in]{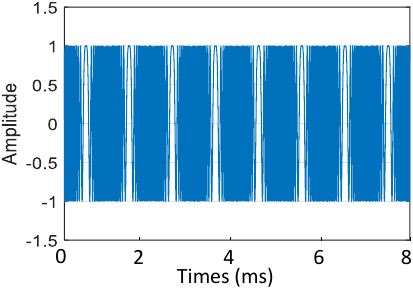}\label{fig:iqsample}}
\hspace{0.2cm}
\subfloat[]{\includegraphics[width=1.54in]{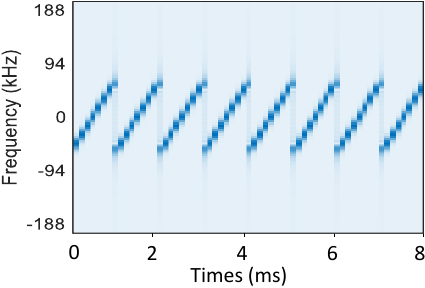}\label{fig:spevtrogram}}
\caption{Eight LoRa preambles. (a) Time domain signal (I branch).  (b) Spectrogram.}
\label{fig:representation}
\end{figure}

LoRa is a well-known low-power wide area network (LPWAN) technology. 
Several repeated preambles at the beginning of the packet are transmitted. These pre-defined sequences are transmitted prior to the payload to enable the receiver to synchronize with the transmitter and they are the same for all the devices. 
LoRa uses chirp spread spectrum (CSS) to modulate the physical layer waveform, whose frequency changes linearly. 
Fig.~\ref{fig:representation}(a) shows the time-domain baseband signal of
the preamble part.
Fig.~\ref{fig:representation}(b) is the spectrogram of the entire preamble part, which indicates how the instantaneous frequency changes over time. 


\subsection{Devices and Dataset}
\begin{figure}[!t]
\centering
\subfloat[]{\includegraphics[width=1.16in]{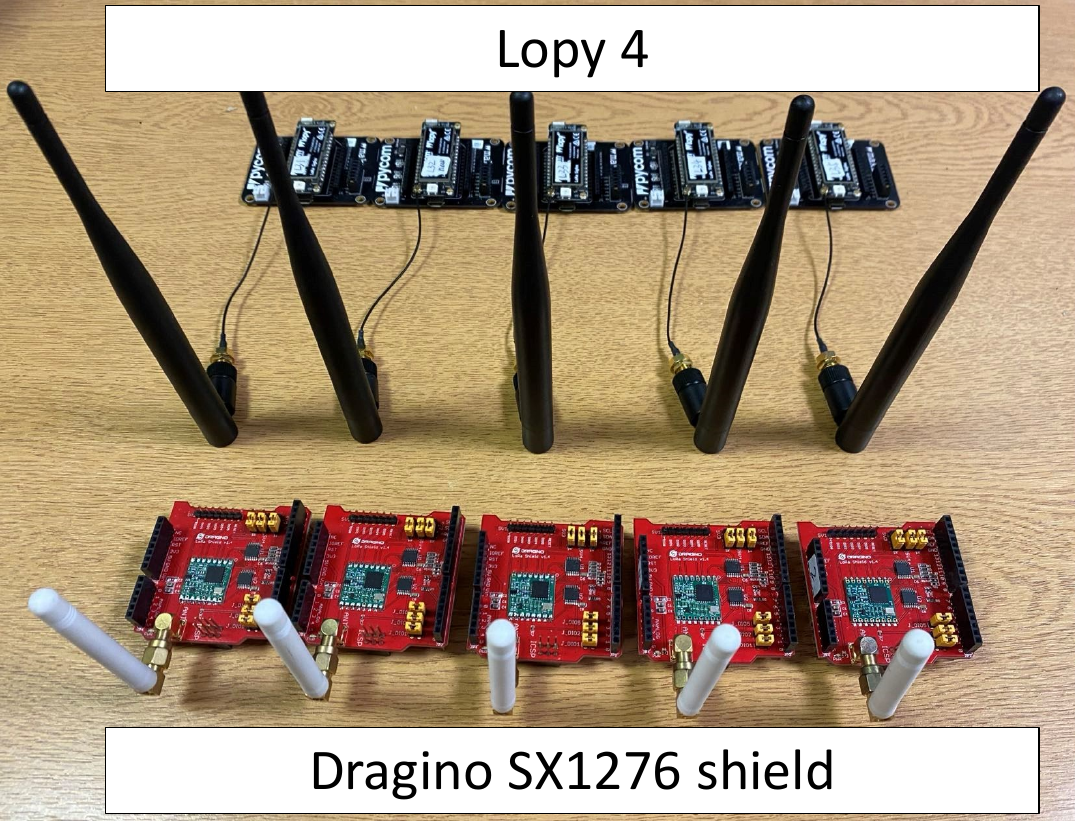}\label{fig:lora}}
\subfloat[]{\includegraphics[width=1.174in]{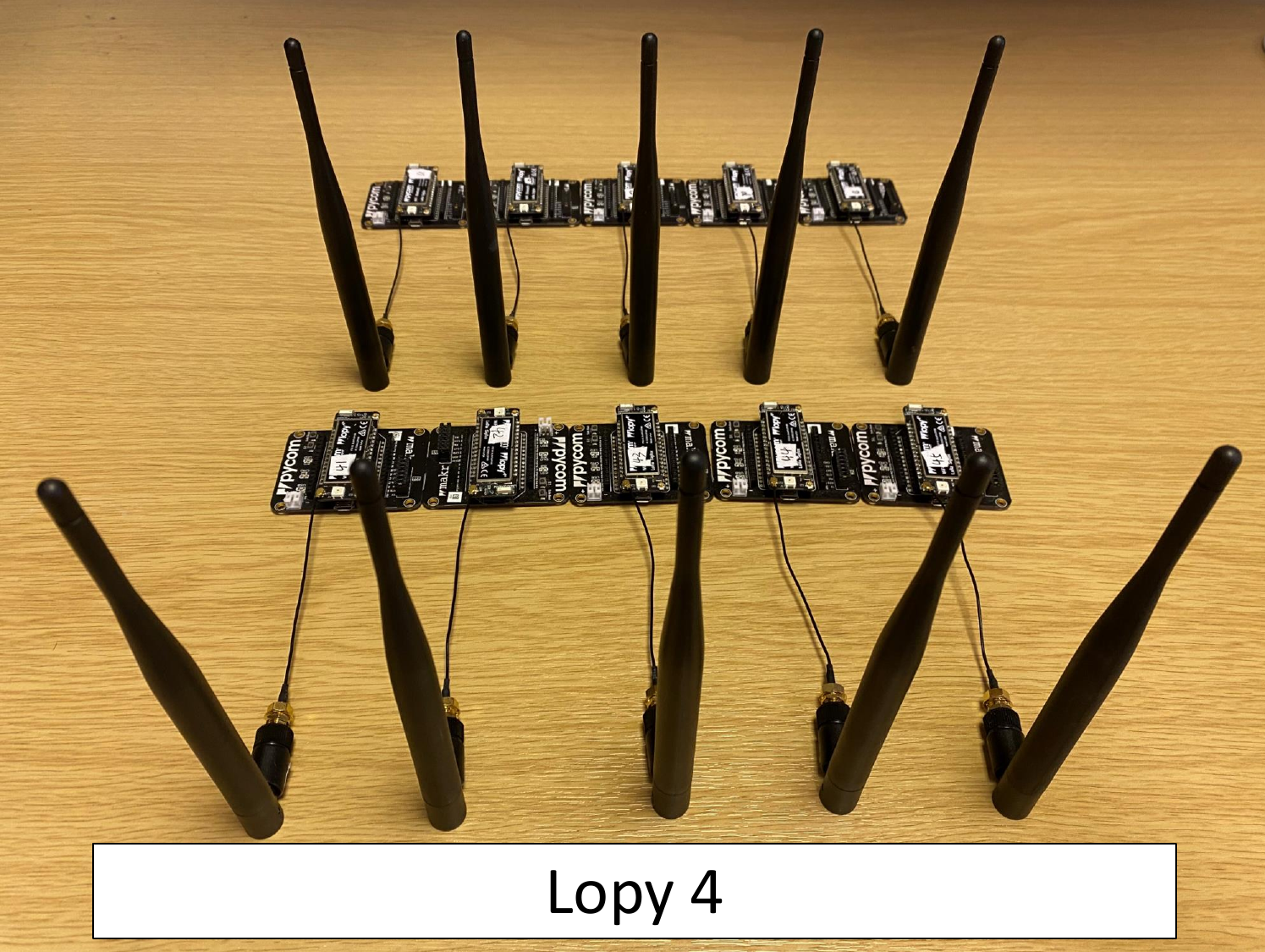}\label{fig:newdevice}}
\subfloat[]{\includegraphics[width=1.18in]{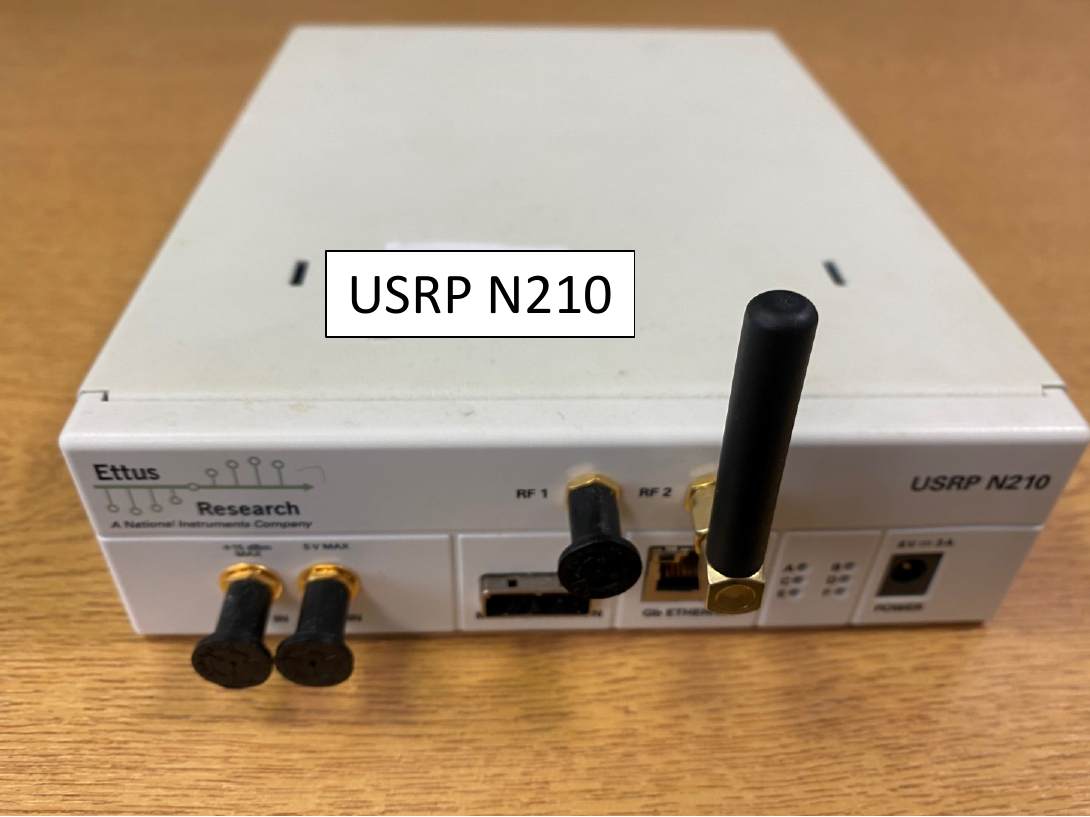}\label{fig:usrp}}
\centering
\caption{Experimental devices. (a) LoRa Device Group 1: D1 - D10. (b) LoRa Device Group 2: D11 - D20. (c) Receiver: USRP N210 SDR platform. }
\label{fig:equip}
\end{figure}

We employed two groups of DUTs. Each group has ten LoRa devices.
As shown in Fig.~\ref{fig:equip}(a), Group 1 consisted of five LoPy4\footnote{https://docs.pycom.io/datasheets/development/lopy4/} boards and five Dragino SX1276\footnote{https://www.dragino.com/products/lora/item/102-lora-shield.html} shields, labelled as D1 to D10. 
Group 2 is depicted in Fig.~\ref{fig:equip}(b), which included ten LoPy4 boards that are designated as D11 to D20.
These devices are all configured with the same parameters, i.e., spreading factor SF7, bandwidth $B=125$~kHz, and carrier frequency $f_c = 868.1$~MHz. The receiver end is a USRP N210 software defined radio (SDR) platform, as shown in Fig.~\ref{fig:equip}(c). Its sampling rate is $f_s = 1$~MHz. The Communications Toolbox Support Package for USRP Radio of MATLAB R2021b is used for accessing the data from USRP. 

We collected a large number of packets from all the LoRa devices in an office environment over five months. 
The LoRa transmitter and the USRP receiver were placed stationary and two meters apart, with line-of-sight (LOS) available. Because of the good channel environment and high SNR, the collected signals are considered to be ideal. 

\subsection{DL-Based RFFI} \label{sec:dlmodel} 
\subsubsection{Signal Representation}
\begin{figure}[!t]
\centering
\subfloat[]{\includegraphics[width=1.2in]{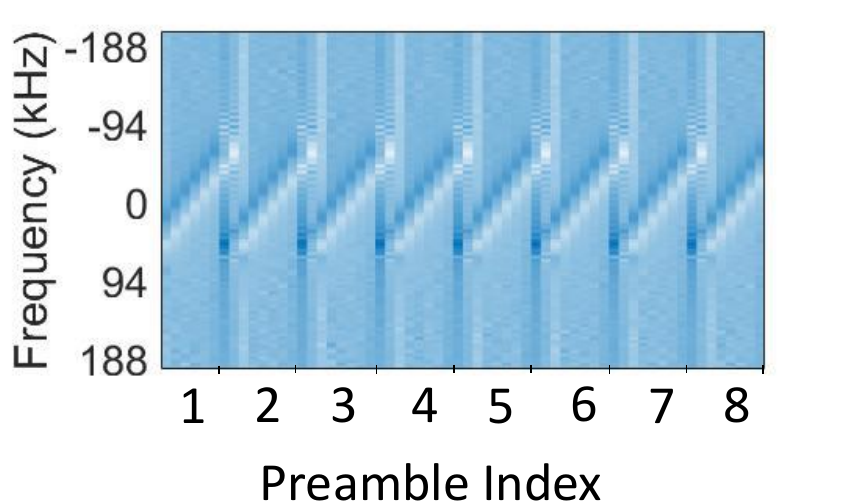}\label{fig:spec_wo_aa}}
\subfloat[]{\includegraphics[width=1.18in]{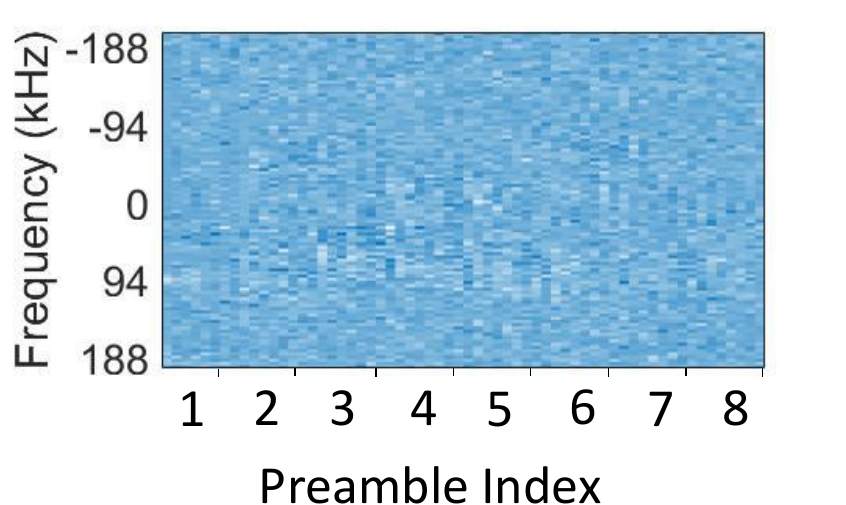}\label{fig:pgdv}}
\subfloat[]{\includegraphics[width=1.18in]{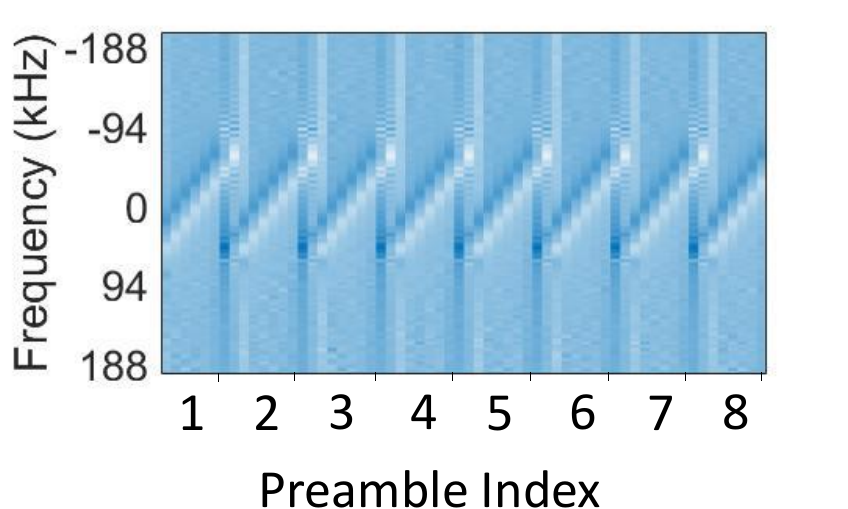}\label{fig:spec_w_aa}}
\caption{(a) Channel-independent spectrogram of clean sample, $x$.  (b) Perturbation calculated by PGD, $v$. (c) Adversarial example, $x^{\prime}$.}
\label{fig:3}
\end{figure}

\mj{While the I/Q samples and the spectrogram shown in Fig.~\ref{fig:representation} can be used for RFFI, they are not robust to channel variations~\cite{towardscalable}. A channel-independent
spectrogram is designed in~\cite{towardscalable} to address the channel variation issue in LoRa-RFFI, which is exemplified in Fig.~\ref{fig:3}(a). 
This paper also uses the channel-independent spectrogram as the signal representation, and consistently, the adversarial attack is directly imposed on the channel-independent spectrogram that serves as the classifier input.}

\subsubsection{Models}\label{sec:dlmodelsetup}
\begin{figure}[t]
\centering
\includegraphics[width=3.4in]{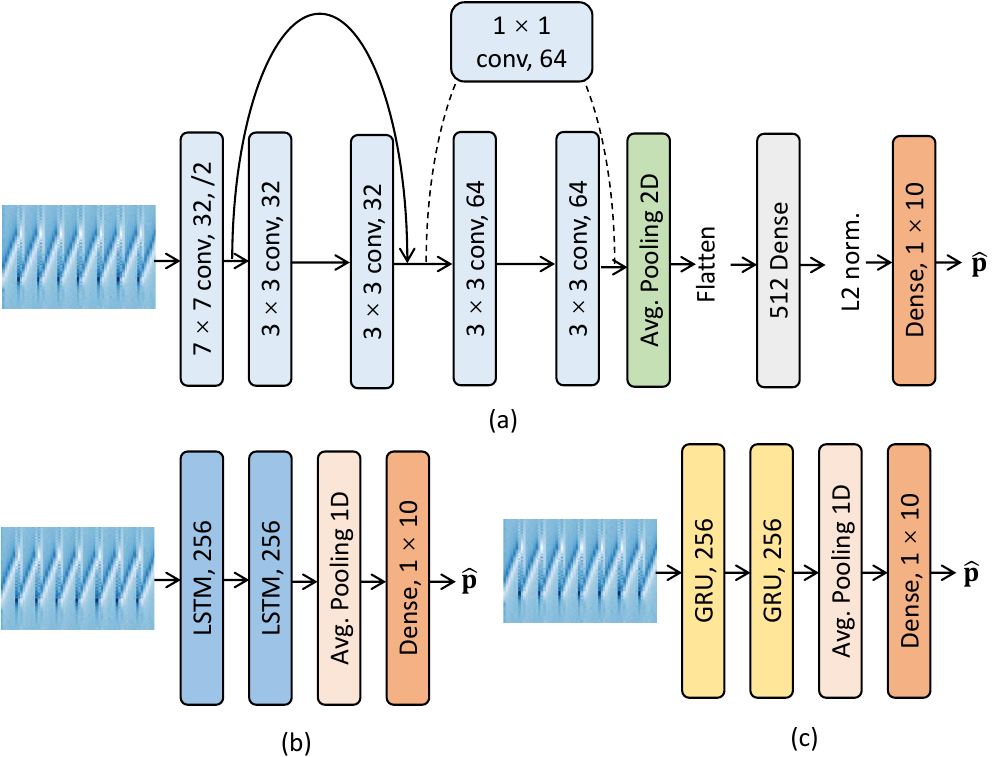}\label{fig:CNN}
\caption{Three DL model architectures. (a) CNN1. (b) LSTM1. (c) GRU1. }
\label{fig:2}
\end{figure}

We studied three widely used DL models for RFFI, namely CNN, LSTM, and GRU. Their architectures are shown in Fig.~\ref{fig:2}.

\begin{itemize}
\item \textbf{CNN} (Fig.~\ref{fig:2}(a)): it consists of one convolutional layer of 32 $7 \times 7$ kernels, followed by two convolutional layers of 32 $3 \times 3$ kernels and two convolutional layers of 64 $3 \times 3$ kernels, one max pooling layer, one 2D average pooling layer, and one dense layer activated by the softmax function. Skip connections are leveraged to improve performance.
\item \textbf{LSTM} (Fig.~\ref{fig:2}(b)): it consists of two 256-unit LSTM layers, one global 1D average pooling layer, and one softmax activated dense layer. 
\item \textbf{GRU} (Fig.~\ref{fig:2}(c)): it is similar to the LSTM (Fig.~\ref{fig:2}(b)), but the LSTM layers are replaced by the GRU layers.
\end{itemize}

The aforementioned models (in Fig.~\ref{fig:2}) are denoted by CNN1, LSTM1, and GRU1, respectively.

\subsection{Training Configuration}
The training of the DL model was carried out on a PC equipped with an NVIDIA GeForce GTX 1660Ti graphic processing unit (GPU). The NNs were implemented with the TensorFlow library. 
All the DL models were trained with the same settings, including validation set ratio, optimizer, learning rate schedule, batch size, and stop condition. $10\%$ of the training data were randomly selected for validation. Adam optimizer was used with an initial learning rate of $1\times10^{-4}$.
The learning rate was reduced by a factor of $0.2$ when the validation loss ceased to decrease for $10$ epochs. 
The batch size was set to $32$. 
The training terminated after the validation loss ceased to improve for $30$ epochs.

\subsection{Classification Results}
\begin{figure}[!t]
\centering
\includegraphics[width=3.4in]{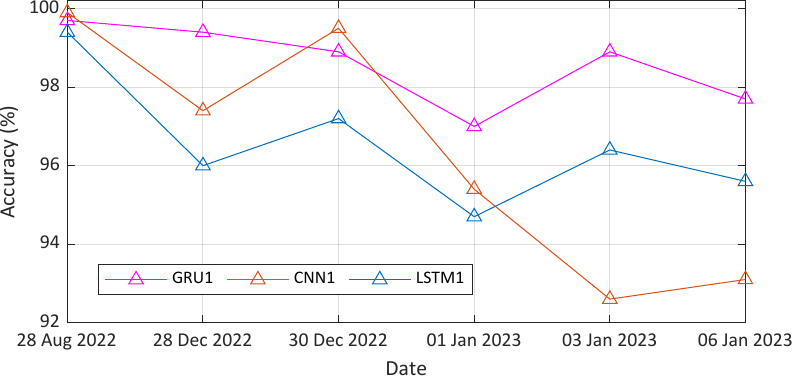}\label{fig:classification_3models}
\caption{Classification accuracy of RFFI system on different dates.}
\label{fig:classification_ori}
\end{figure}

We trained the above three DL-based RFFI systems using datasets collected from Group~1 LoRa devices over four days. $3.6 \times 10^4$ packets from each device were collected.
The trained models were tested on the various datasets collected from different days. Their classification performances are shown in Fig.~\ref{fig:classification_ori}.
The results indicate that all the trained CNN1/LSTM1/GRU1-based RFFI systems can achieve above 92\% classification accuracy, which demonstrated that our trained RFFI models have good generalization.

\section{Experimental Evaluation of Perturbation Generation Methods}\label{sec:experi_eva}
In this section, we evaluated the attack performance of FGSM, PGD, and UAP on RFFI systems. The DL models of the victim RFFI systems are the same as those described in Section~\ref{sec:dlmodel}. The dataset was collected from Group 1 LoRa devices. We used additive white Gaussian noise (AWGN) as the baseline to evaluate the adversarial attack performance on the RFFI system operating in the presence of natural noise.
Each evaluation has a specific adversary setup per attack strategy, which will be explained in the following sections.

\subsection{Evaluation of FGSM and PGD Under White-Box Attacks}\label{sec:white-box} 

\subsubsection{Setup}
The adversary utilizes exactly the same setup as the victim, which includes the same dataset and DL models, and generates perturbations. We used a prefix-suffix notation ``\textit{A}-\textit{B}" to denote a perturbation generated by the method \textit{A} on a DL model \textit{B}. For example, ``FGSM-CNN1'' refers to an attack on a CNN1-based RFFI system using the perturbation generated by FGSM. Our implementations of FGSM and PGD are based on the standardized reference implementations of adversarial attacks provided by CleverHans.v3.1.0~\cite{papernot2018cleverhans}. It is a software library used for benchmarking machine learning systems' vulnerability to adversarial examples.

\subsubsection {Results - Non-targeted Attacks with Different PSRs} \label{sec:diffPSR}
The perturbation generated by PGD ($\emph{PSR}=-30$~dB) is shown in Fig.~\ref{fig:3}(b).
The perturbation is superimposed onto the clean sample shown in Fig.~\ref{fig:3}(a) to produce the adversarial example shown in Fig.~\ref{fig:3}(c). It can be observed that the clean sample and the adversarial example are visually indistinguishable, which also indicates that the adversarial attack did not cause too much degradation to the signal.

\begin{figure}[!t]
\centering
\subfloat[]{\includegraphics[width=1.7in]{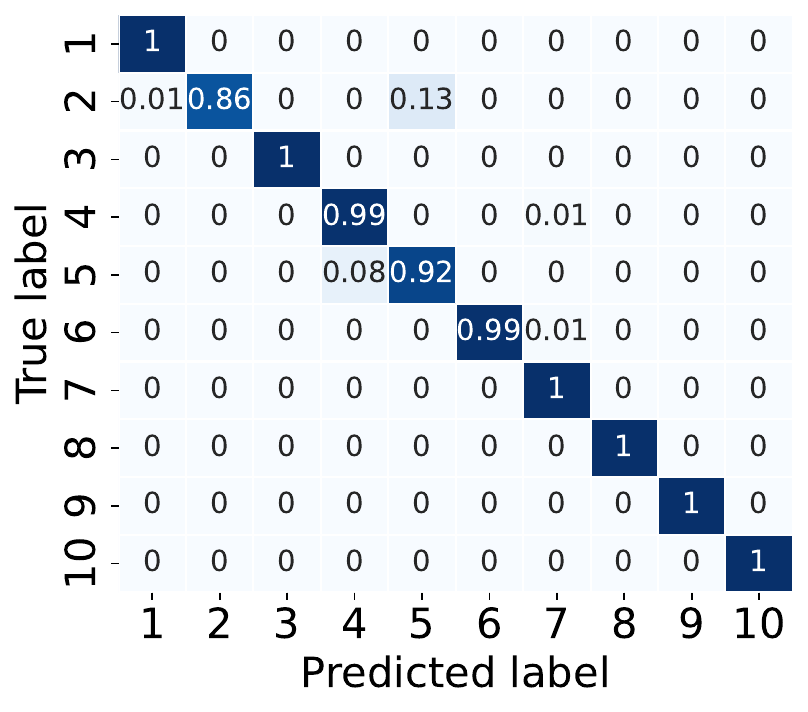}\label{fig:12}}
\hspace{0.1cm}
\subfloat[]{\includegraphics[width=1.7in]{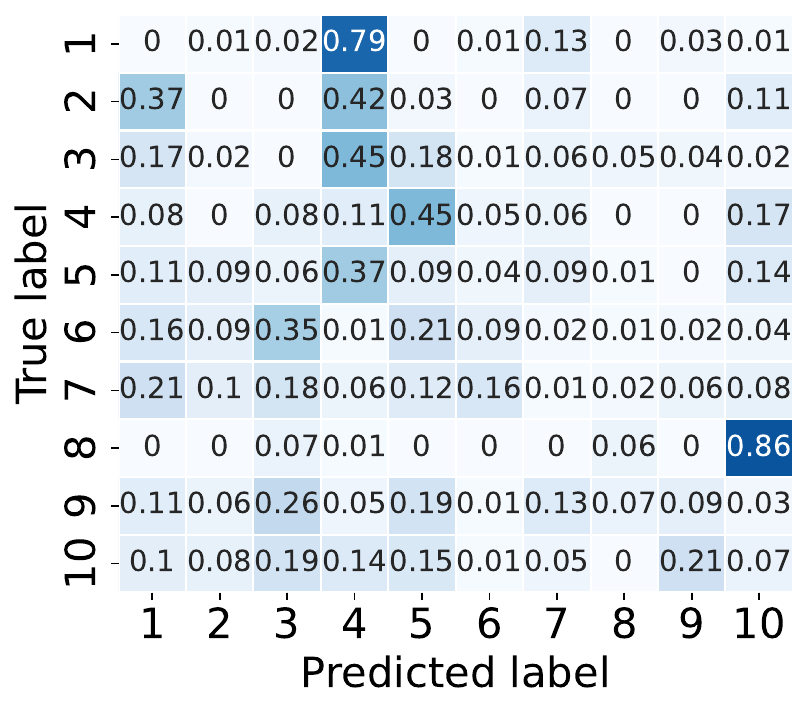}\label{fig:22}}
\\
\subfloat[]{\includegraphics[width=1.7in]{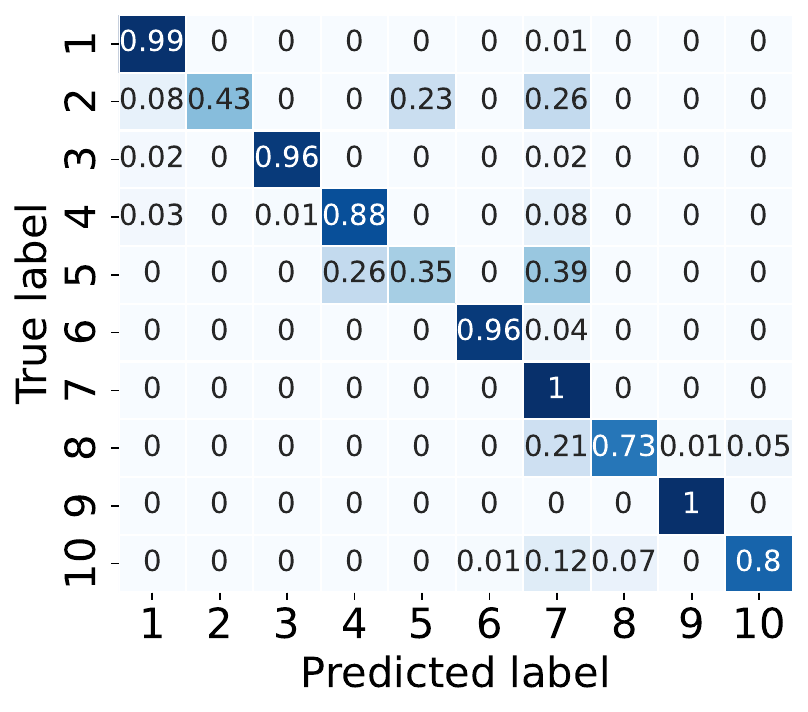}\label{fig:13}}
\hspace{0.1cm}
\subfloat[]{\includegraphics[width=1.7in]{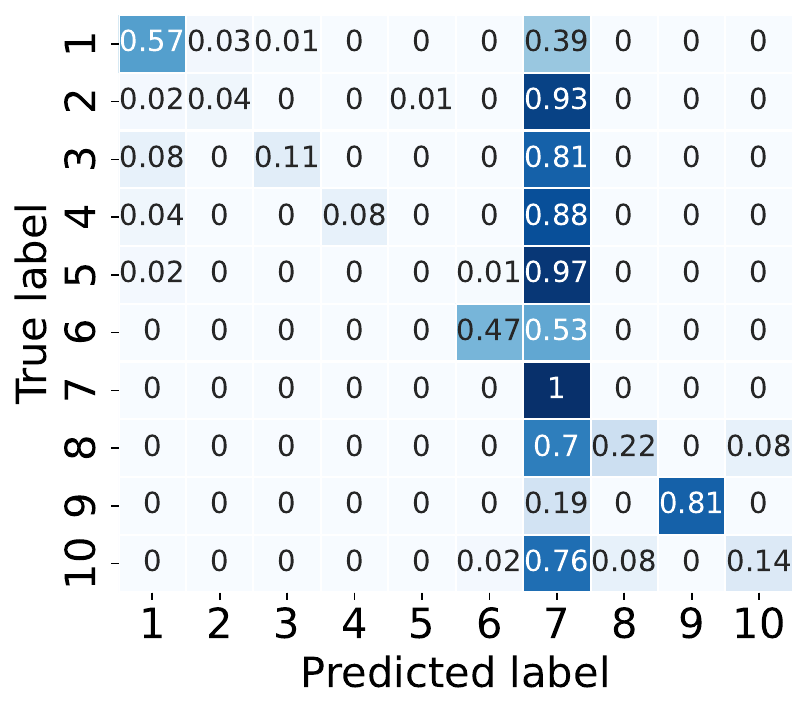}\label{fig:14}}
\caption{Confusion matrix. (a) RFFI without adversarial attack. (b) CNN1-PGD with non-targeted attacks. $\emph{PSR}=-30$~dB. (c) CNN1-PGD with targeted attacks. $\emph{PSR}=-30$~dB. (d) CNN1-PGD with targeted attacks. $\emph{PSR}=-25$~dB. }
\label{fig:15}
\end{figure}

Fig.~\ref{fig:15}(a) and Fig.~\ref{fig:15}(b) show the confusion matrix of the CNN1-based RFFI without attack and that of a non-targeted attack using perturbations produced by PGD when $\emph{PSR}=-$30~dB, respectively.
The overall accuracy reduces from $97.6\%$ to $5.2\%$ upon attack. As the attack is untargeted, the incorrect predictions disperse across all labels.

\begin{figure}[t]
\centering
\includegraphics[width =3.4in]{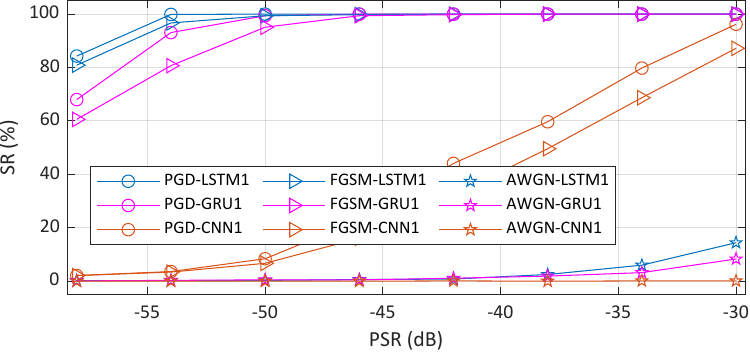}
\caption{SR of FGSM and PGD with non-targeted attacks against PSRs. 
}
\label{fig:7}
\end{figure}

Fig.~\ref{fig:7} compares the performance of non-targeted attacks when the PSR varied from $-58$~dB to $-30$~dB.
DL-based RFFI systems suffer more severe misclassifications at higher PSR. Specifically, when the PSR is $-54$~dB, the SRs of CNN1-FGSM RFFI and CNN1-PGD RFFI reach 87.2\% and 96.2\%, respectively. 
The iterative PGD is found to be more effective than the one-step FGSM.
The SRs of all the adversarial attacks increases with increasing PSR. It should be noted that raising the PSR requires more transmit power, making the adversarial attack less stealthy.

Fig.~\ref{fig:7} shows that there is hardly any effect on the performance of these RFFI classifiers when the same power of AWGN as the adversarial perturbation is injected into the input signal. Although RFFI classifiers are robust against AWGN, their prediction accuracies degrade significantly in the presence of adversarial examples.

\subsubsection{Results - Targeted Attacks} \label{sec:pgd-target}
\begin{figure}[!t]
\centering
\includegraphics[width =3.4in]{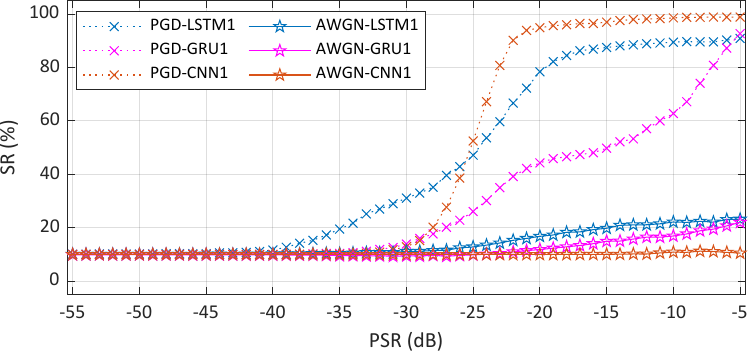}
\caption{SR of PGD with targeted attacks against PSRs.}
\label{fig:psr_target}
\end{figure}

PGD was utilized to implement targeted attacks. We arbitrarily selected D7 as the target DUT.
Fig.~\ref{fig:15}(c) shows the confusion matrix of PGD-CNN1 with the targeted attack at $\emph{PSR}=-30$~dB, where 21.3\% of the devices are misclassified as D7. 
When the PSR increases to $-25$~dB, 71.6\% of the devices are misclassified as D7, as shown in Fig.~\ref{fig:15}(d). 
The SR of the targeted attack can be further increased with higher PSR as shown in Fig.~\ref{fig:psr_target}. 
An increasing number of devices were misclassified as the target (D7) with increasing PSR. 
When the PSR is $-5$~dB in the CNN1-based RFFI, almost all packets ($\sim$98.9\%) are incorrectly identified as being transmitted from the target device.
By comparing Fig.~\ref{fig:7} and Fig.~\ref{fig:psr_target}, we can observe that a higher PSR is required in a targeted attack to achieve the same SR as a non-targeted attack. This is expected as it is harder to cause a model to output a specific wrong label rather than any incorrect label.

\begin{figure*}[!t]
\centering
\subfloat[]{\includegraphics[width=2.0in]{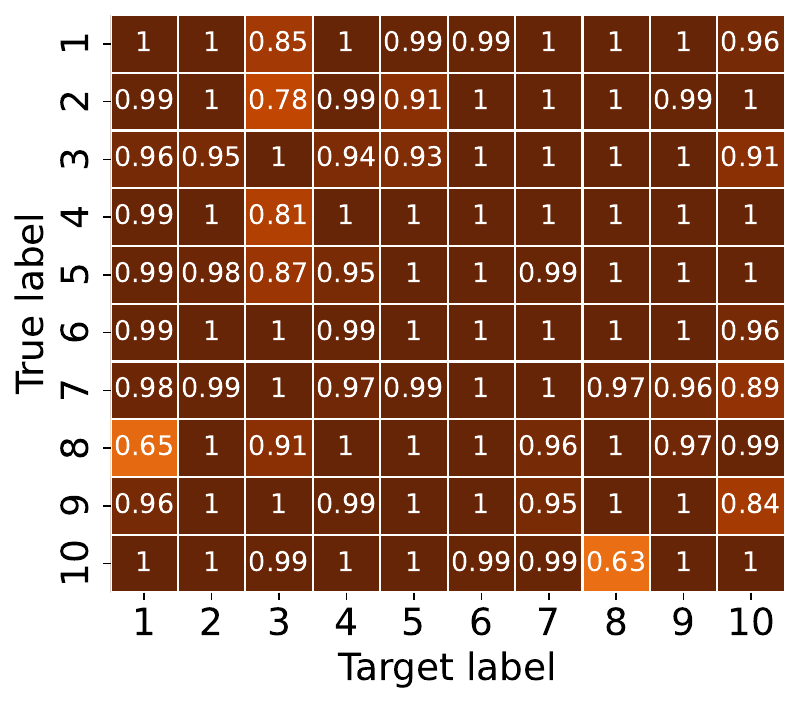}\label{fig:16}} \hspace{0.01in}
\subfloat[]{\includegraphics[width=2.0in]{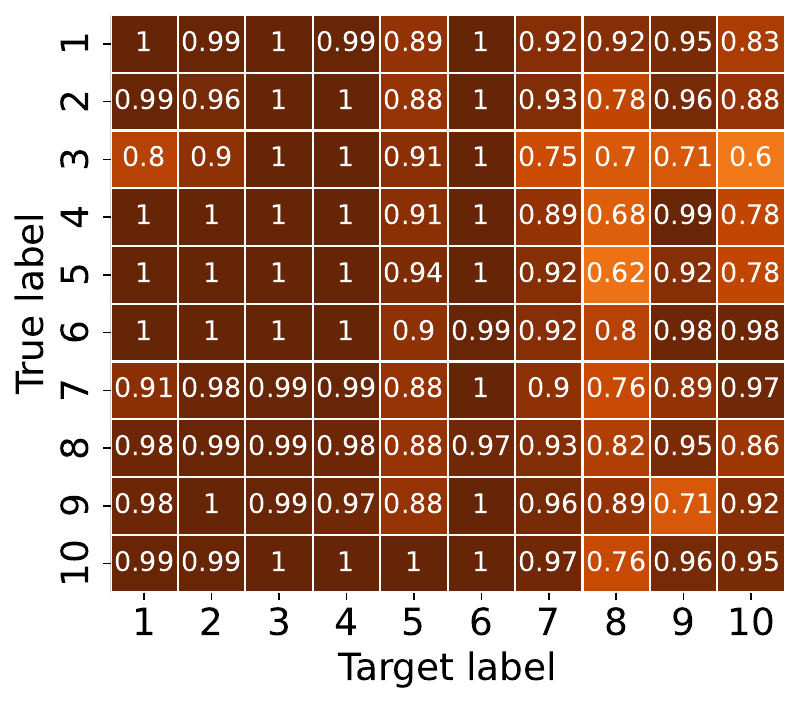}\label{fig:17}}
\hspace{0.01in}
\subfloat[]{\includegraphics[width=2.0in]{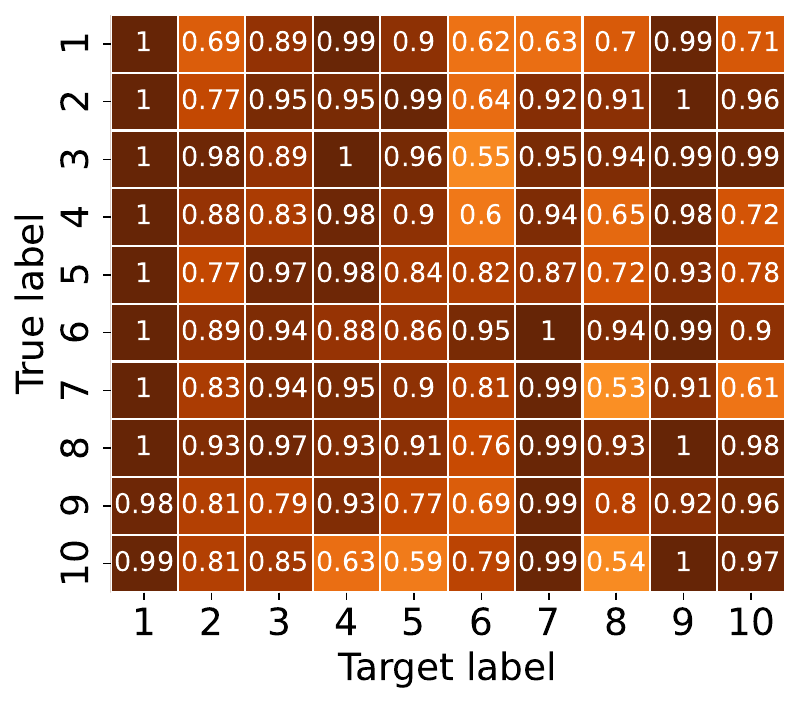}\label{fig:18}}
\centering
\caption{Targeted attacks by PGD. $\emph{PSR}=-5$~dB. (a) CNN1-based RFFI. (b) LSTM1-based RFFI. and (c) GRU1-based RFFI.}
\label{fig:99}
\end{figure*}

\mj{To further investigate the performance of targeted attacks of different selected targets, we set each device as the target one by one. The results are shown in Fig.~\ref{fig:99}. Different from a confusion matrix, each matrix element in Fig.~\ref{fig:99} represents the ratio of devices identified as the selected target.}
For example, 85\% of the packets from D1 are identified as originating from D3 in CNN1-based RFFI when the selected target is D3 (Fig.~\ref{fig:99}(a)). Apparently, PGD can cause the victim model to indiscriminately misclassify most devices as the selected target. It suggests that PGD allows the adversary to masquerade as any legal device that can be recognized by an RFFI system, including the most privileged user device, to gain successful authentication. 

FGSM and PGD do not show advantages in the context of grey/black-box attacks~\cite{discretization, bestblack, Brunner_2019_ICCV}. As each perturbation obtained by these two methods is computed based on its respective input, which cannot be satisfied under grey/black-box attacks. Therefore, FGSM and PGD under grey/black-box attacks have not been discussed.

\subsection{Evaluation of Universal Adversarial Perturbation}\label{sec:UAP} 
\subsubsection{Setup}\label{sec:UAPsetup} 
We evaluated the performance of UAP under white-box, grey-box, and black-box settings.
In the white-box attack, the adversary will use the same DL model and dataset as the victim system to generate the perturbations by UAP method. In the grey/black-box attacks, the adversary will have to rely on a surrogate dataset/model to produce the perturbations. This is feasible thanks to the transferability~\cite{papernot2016transferability} of UAP. Transferability here refers to the portability of an adversarial perturbation to more than one network model. By combining the generalizable neural network with the cross-entropy loss, the surrogate model can be trained to generate the UAPs that are effective on the victim model.

In this subsection, we assume the surrogate devices are the same as the victim (Group 1 LoRa devices).
However, the surrogate dataset is different from the victim dataset, as it was collected separately. To ensure the surrogate model exhibits high generalization, its training dataset also reached $3.6 \times 10^4$ packets from each device.

The surrogate models used are delineated as follows:
\begin{itemize}    
    \item The surrogate \textbf{CNN} model has two fewer layers than the victim CNN model. More specifically, the second and third convolutional layers in Fig.~\ref{fig:2}(a) were removed. 
    \item The surrogate \textbf{LSTM} model utilizes one 128-unit LSTM layer to replace the first 256-unit LSTM layer in Fig.~\ref{fig:2}(b). 
    \item The surrogate \textbf{GRU} model utilizes one 128-unit GRU layer to replace the first 256-unit GRU layer in Fig.~\ref{fig:2}(c). 
\end{itemize}

These surrogate models are denoted by CNN2, LSTM2, and GRU2, respectively.


\subsubsection{Results}  

\begin{figure}[!t]
\centering
\subfloat[]{\includegraphics[width=3.4in]{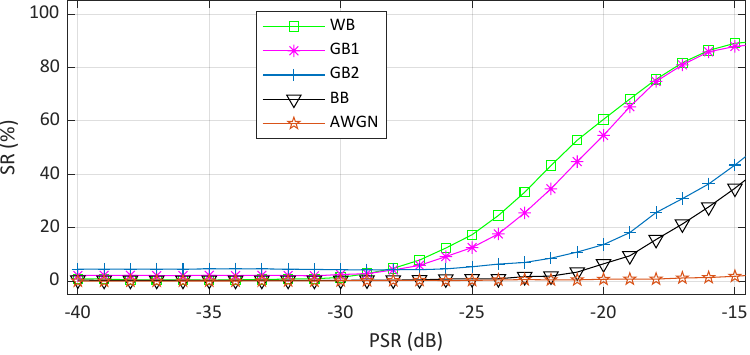}\label{fig:CNN-UAP}}

\subfloat[]{\includegraphics[width=3.4in]{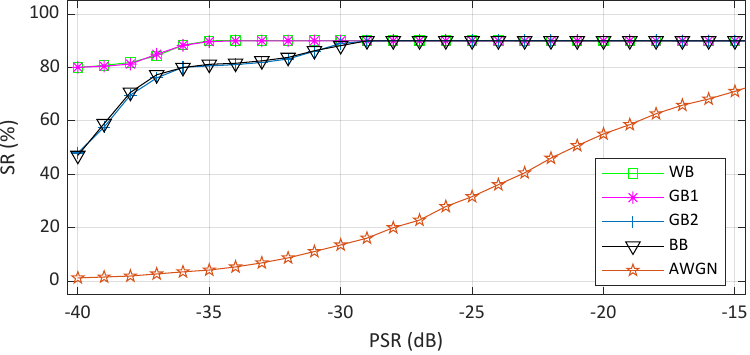}\label{fig:LSTM-UAP}}\hspace{0.1cm}

\subfloat[]{\includegraphics[width=3.4in]{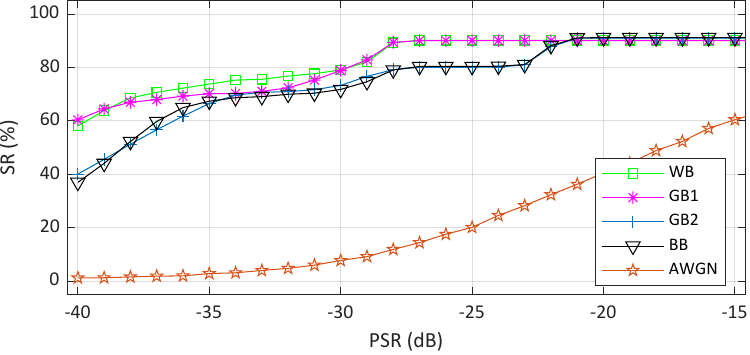}\label{fig:GRU-UAP}}
\centering
\caption{SR of UAP under different attack scenarios against PSRs. (a) CNN1-based RFFI, (b) LSTM1-based RFFI, and (c) GRU1-based RFFI.}
\label{fig:UAPdiffmodels}
\end{figure}

The SRs of attacking CNN1, LSTM1, and GRU1-based RFFI by UAP are shown in Fig.~\ref{fig:UAPdiffmodels}, with PSR varying from $-40$~dB to $-15$~dB.
At a PSR of $-30$~dB, the SR of the attack on LSTM1-based RFFI is 88.2\% when the victim's dataset and DL model are unknown, and it is still higher than AWGN.

The SR of UAP for the grey/black-box attacks on the same victim model with the same PSR is lower than that of the white-box attack. Unlike the white-box attack, which can be specially crafted based on the decision boundaries of the actual DL model of the victim system, the grey/black-box attacks can only approximate the actual decision boundaries of the victim classifier through UAP transferability. 
The SR of UAP converges to 90\% and no longer rises with increasing PSR, which can be attributed to the dominant characteristic of the UAP~\cite{Zhang_2021_ICCV, Moosavi-Dezfooli_2017_CVPR, uapdom}. All the packets will be identified as transmitted from a specific device when PSR is high, which means the packets from that device are correctly classified.

In addition, the results indicate that the performance of GB1 is close to WB. It can be attributed to the high generalization of DL models and the similarity of surrogate dataset to the victim dataset. Both the surrogate and victim models have good generalizability and are fed with the dataset collected from LoRa Device Group 1, which results in their decision boundaries being close.

Compared to Fig.~\ref{fig:7}, UAP is less effective than FGSM/PGD for the same PSR.
The SR of FGSM/PGD is close to 100\% on LSTM1 and GRU1 when $\emph{PSR}=-40$~dB. However, the SR of UAP is about 80\% on LSTM1 and 60\% on GRU1 under the same white-box setting. Being sample-agnostic, UAP does not have the opportunity like FGSM/PGD to perform sample-specific optimization. Hence, a higher PSR is required for UAP to achieve the same SR as FGSM/PGD.

\section{Adversarial Attacks To LoRa-RFFI Under Practical Scenarios}\label{sec:practical_tests}

\mj{Section~\ref{sec:experi_eva} presented adversarial attacks using FGSM, PGD, and UAP. However, in realistic wireless contexts, FGSM and PGD are not practical as they need to generate a unique perturbation for a specific sample/packet. Therefore, in this section, we explored UAP for adversarial attacks considering factors related to wireless communications. In particular, we consider the effects of DL models, the attack time, surrogate devices, and real-time attacks separately. We then combine all the above aspects to investigate the adversarial attack in more practical settings. 
Table~\ref{tab:label_test} summarizes the different evaluation and their setups.
\begin{table*}
\caption{The summary of experimental setup. 
The training dataset for the surrogate models and victim models without * are from Group 1, collected on 28th Aug., 2nd, 7th, 15th Sep. 2022. 
The training dataset for the victim models marked with * are from Group 2, collected on 25th and 28th Jan. 2023.}
  \resizebox{\textwidth}{!}{%
  \begin{threeparttable}
\begin{tabular}{|l|l|l|l|l|l|}
\hline
Section & Purpose                                                    & Surrogate Model                                                                 & Victim Model                                                                     & Surrogate Dateset          & Victim Dataset                                                                                                                                                                                    \\ \hline\bigstrut
VII-A   & The   effect of attack across different DL models          & \begin{tabular}[c]{@{}l@{}}CNN1, LSTM1, GRU1, \\ CNN2, LSTM2, GRU2\end{tabular} & \begin{tabular}[c]{@{}l@{}}CNN1, LSTM1, GRU1, \\ CNN2, LSTM2, GRU2\end{tabular}  & $\mathcal{D}^{d1}_{G1}$ & $\hat{\mathcal{D}}^{d1}_{G1}$                                                                                                                                                                        \\ \hline\bigstrut
VII-B   & The   effect of attack across different days               & CNN1, LSTM1, GRU1                                                               & CNN1, LSTM1, GRU1                                                                & $\mathcal{D}^{d1}_{G1}$ & \begin{tabular}[c]{@{}l@{}}$\hat{\mathcal{D}}^{d1}_{G1}$, $\hat{\mathcal{D}}^{d2}_{G1}$, $\hat{\mathcal{D}}^{d3}_{G1}$,\\ $\hat{\mathcal{D}}^{d4}_{G1}$, $\hat{\mathcal{D}}^{d5}_{G1}$, $\hat{\mathcal{D}}^{d6}_{G1}$\end{tabular} \\ \hline\bigstrut
VII-C   & The   effect of attack across different devices            & \begin{tabular}[c]{@{}l@{}}CNN1, LSTM1, GRU1, \\ CNN2, LSTM2, GRU2\end{tabular} & \begin{tabular}[c]{@{}l@{}}CNN1, LSTM1, GRU1, \\ CNN1*, LSTM1*,  GRU1*\end{tabular} & $\mathcal{D}^{d1}_{G1}$ & $\hat{\mathcal{D}}^{d1}_{G1}$, $\hat{\mathcal{D}}^{\hat{d1}}_{G2}$                                                                                                                                               \\ \hline\bigstrut
VII-D   & The   effect of real time attack                           & CNN2, LSTM2, GRU2                                                               & CNN1, LSTM1, GRU1                                                                & $\mathcal{D}^{d1}_{G1}$ & $\hat{\mathcal{D}}^{d1}_{G1}, \hat{\mathcal{D}}^{d2}_{G1}$                                                                                                                                                                        \\ \hline\bigstrut
VII-E   & The   effect of attack involving all the mentioned aspects & CNN2, LSTM2, GRU1                                                               & CNN1*, LSTM1*, GRU1*                                                                & $\mathcal{D}^{d1}_{G1}$ & $\hat{\mathcal{D}}^{\hat{d1}}_{G2}$,  $\hat{\mathcal{D}}^{\hat{d2}}_{G2}$,  $\hat{\mathcal{D}}^{\hat{d3}}_{G2}$                                                                                                                    \\ \hline
\end{tabular}
\begin{tablenotes}
 \item[] Symbol explanation: $\mathcal{D}$ denotes Surrogate Dataset to calculate perturbation, $\hat{\mathcal{D}}$ denotes Victim Dataset being add perturbation, ${G1}/G2$ denotes Group 1/2 devices, ${dx}$ denotes day $x$
\end{tablenotes}
\end{threeparttable}
}
  \label{tab:label_test}
\end{table*}}

\subsection{Cross-Model Evaluation}\label{sec:cross-UAP}
\mj{The surrogate models used for the study of UAP attacks under grey/black-box settings in Section~\ref{sec:UAP} differ from the corresponding victim model by only one or two layers. In practice, the adversary may not even know the type of DL model adopted. We explore the effect of UAPs generated by surrogate models that are structurally deviated from the victim DL models in this section.}
\subsubsection{Setup}\label{sec:UAPsetup1} 

We generated the UAPs with $\mathcal{D}_{G1}^{d1}$ using six different DL models in order to evaluate their transferability across structurally deviated models. 
The six surrogate models used for this evaluation include CNN1, CNN2, LSTM1, LSTM2, GRU1, and GRU2, which has been introduced in Section~\ref{sec:dlmodelsetup} and Section~\ref{sec:UAPsetup}.

\subsubsection{Results}
As shown in Table~\ref{tab:frindiffmodel}, the overall results indicate reasonably good transferability for most of the heterogeneous surrogate-victim model pairs experimented, especially between LSTM and GRU. 
For example, the UAP generated by LSTM1 has a SR of 90\% on LSTM2, 90\% on GRU2, and 90.1\% on GRU1. UAP transferability between CNN and LSTM/GRU is poorer. The worst SR is the attack on CNN2 using LSTM2 as a surrogate model for UAP generation. 
\begin{table}[!t]
  \centering
  \caption{SR of UAP in Cross-model Scenarios. }
	\scalebox{0.85}{
        \begin{tabular}{|l|l|l|l|l|l|l|}
    \hline
    \diagbox{Surrogate}{Victim} & CNN1 & CNN2 & LSTM1 & LSTM2 & GRU1 & GRU2 \bigstrut\\
    \hline
    CNN1& \textbf{90\%} & 57.4\%  & 81.2\%    & 80.2\%  & 74.2\%  & 64.6\% \bigstrut\\
    \hline
    CNN2 & 52.3\%  & \textbf{89.9\%} & 86.1\%    & 80.4\%  & 74.5\%    & 68.5\% \bigstrut\\
    \hline
    LSTM1 & 68.4\%  & 66.8\% & \textbf{90\%} & 90\%  & 90\%  & 90.1\%
    \bigstrut\\
    \hline
    LSTM2 & 78.5\%  & 49.4\%  & 90\%    & \textbf{90\%} & 90\%   & 87.2\% \bigstrut\\
    \hline
    GRU1 & 78\%    & 72.9\%  & 90\%  & 90\%  & \textbf{90\%} & 90\% 
    \bigstrut\\
    \hline
    GRU2 & 86.8\%    & 87.1\%  & 90.1\%    & 90\%  & 90\%    & \textbf{90\%} \bigstrut\\
    \hline

    \end{tabular}}
  \label{tab:frindiffmodel}%
\end{table}%

The observed variations in SRs for different surrogate-victim model pairs in Table~\ref{tab:frindiffmodel} can be attributed to the extent of the architectural difference between the surrogate and victim models. LSTM and GRU are variants of recurrent neural networks (RNNs). The difference in their network architectures is not as significant as the architectural difference between RNN and CNN. The results indicate a high probability that UAP can succeed in impersonating any device without having to know the victim DL model's architecture used by the RFFI system, especially the surrogate model used for the perturbation generation belongs to the same major NN architecture.

\subsection{Cross-Day Evaluation}\label{sec:cross-time} 


\subsubsection{Setup}
We generated UAPs with $\mathcal{D}_{G1}^{d1}$. 
The UAPs were then superimposed onto other test datasets from Group 1, which were gathered on the following dates: 28$^{th}$ and 30$^{th}$ December 2022 as well as 1$^{st}$, 3$^{rd}$, and 6$^{th}$ January 2023. These five datasets are denoted as $\hat{\mathcal{D}}_{G1}^{d2}$ to $\hat{\mathcal{D}}_{G1}^{d6}$.

The victim and surrogate DL models include CNN1, LSTM1, and GRU1. We chose the UAPs with $\emph{PSR}=-16$~dB to $\emph{PSR}=-14$~dB for CNN1; $\emph{PSR}=-39$~dB to $\emph{PSR}=-37$~dB for LSTM1; $\emph{PSR}=-32$~dB to $\emph{PSR}=-30$~dB for GRU1. 

\subsubsection{Result}
\begin{figure}[!t]
    \centering
    \subfloat[]{\includegraphics[width=3.4in]{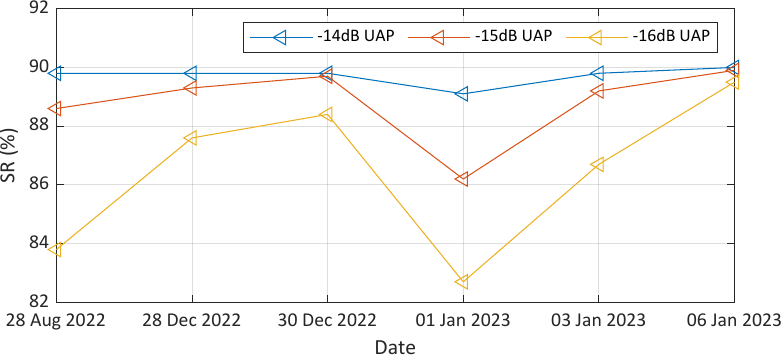}\label{fig:CNNDIFPER}}
    
    \subfloat[]{\includegraphics[width=3.4in]{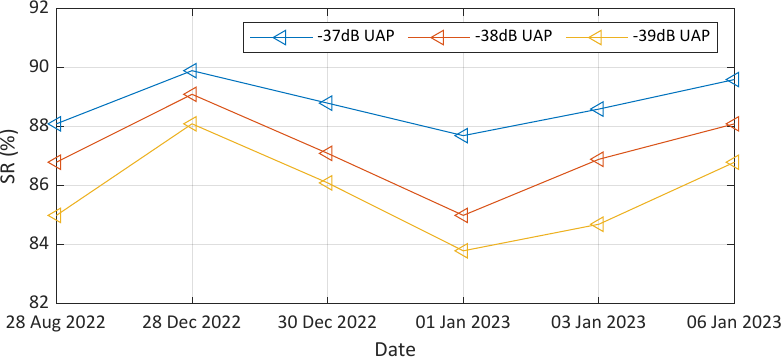}\label{fig:LSTMDIFPER}}
    
    \subfloat[]{\includegraphics[width=3.4in]{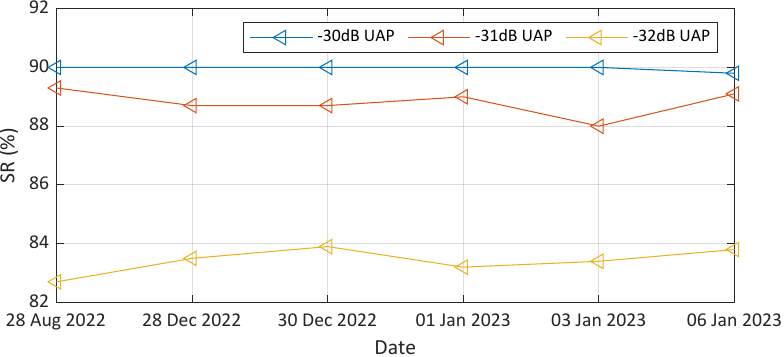}\label{fig:GRUDIFPER}}
    \centering
    \caption{SR of UAP on different dates. (a) CNN1-based RFFI. (b) LSTM1-based RFFI. and (c) GRU1-based RFFI.}
    \label{fig:uap_timeperiod}
\end{figure}

The SRs of UAP on DL-based RFFI on different days are shown in Fig.~\ref{fig:uap_timeperiod}.
From Fig.~\ref{fig:uap_timeperiod}(a), the SR of UAP ($\emph{PSR}=-16$~dB) attack mounted on 28$^{th}$ August 2022 on CNN1 is 83.8\%, and the SRs of the same attack on the following days fluctuate around 83.8\%. The same trend is also observed with other PSRs, and attacks on both LSTM1-based RFFI (Fig.~\ref{fig:uap_timeperiod}(b)) and GRU1-based RFFI (Fig.~\ref{fig:uap_timeperiod}(c)).
\mj{The fluctuations indicate that the dataset itself influences the SR. At the same time, the fact that the SR of the UAP does not decrease significantly over time suggests that the adversary does not need to frequently retrain the surrogate model or update the UAP to maintain attack effectiveness.}

\subsection{Cross-Device Evaluation}\label{sec:cross-device}
The previous discussion is based on the assumption that the adversary obtains the surrogate dataset from the victim devices to be attacked. In practice, this is possible because the adversary could eavesdrop on the wireless transmissions from the victim's devices. However, it would take a huge amount of time and the quality of the dataset may not be good, because the adversary has no control over the victim system, including the transmission interval, SNR, etc.

The adversary can build a surrogate system by using a hardware testbed with the same wireless protocol and collect the dataset on its own.
This section will explore whether the UAP generated by the surrogate devices will be applicable to the victim system.


\subsubsection{Setup}\label{sec:cross-devicesetup}
We changed the victim RFFI system to the Group 2 LoRa devices, i.e., ten LoPy4 boards shown in Fig.~\ref{fig:equip}(b). 
The training dataset was collected between January 25$^{th}$ and January 28$^{th}$, 2023. The same NN architecture as presented in Section~\ref{sec:dlmodelsetup} was employed.
The trained models were then denoted CNN1*, LSTM1*, and GRU1*, respectively. The perturbations applied under different levels of assumptions are explained as follows.



\begin{itemize}
    \item \textbf{UAP-Same-Device:} refers to the UAP in an ideal scenario, both of the surrogate and victim datasets are from Group 1, and the surrogate and victim models are the same.
    \item \textbf{UAP-Diff-Device:} the surrogate devices used by the adversary are from Group 1, which are different from the victim system. The surrogate models are the same as the victim RFFI DL models.    
    \item \textbf{UAP-Diff-Device\&Net:} the surrogate devices used by the adversary are from Group 1. The surrogate models are different from the victim DL models. The UAP generated by the CNN2/LSTM2/GRU2-based RFFI system will be used to attack the CNN1*/LSTM1*/GRU1*-based RFFI system correspondingly.    

\end{itemize}
We also compared the SR against the misclassification due to AWGN.

\subsubsection{Result}
\begin{figure}[!t]
\centering
\subfloat[]{\includegraphics[width=3.4in]{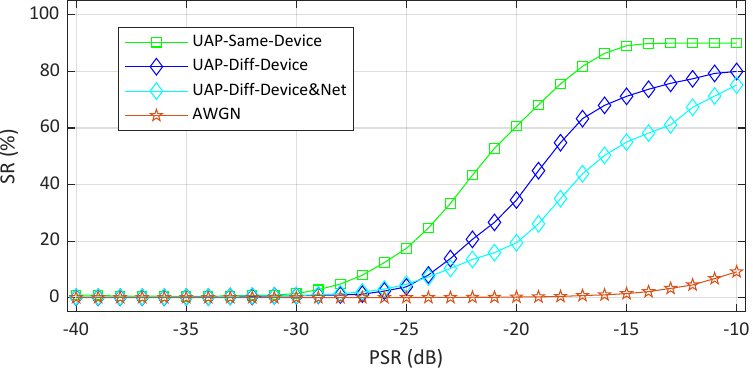}\label{fig:CNNDIFDEC}}

\subfloat[]{\includegraphics[width=3.4in]{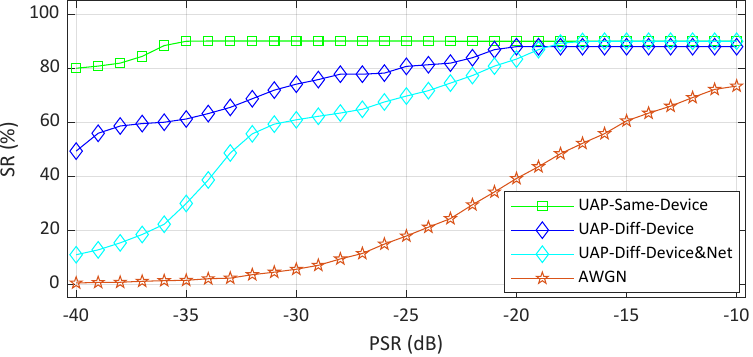}\label{fig:LSTMDIFDEC}}

\subfloat[]{\includegraphics[width=3.4in]{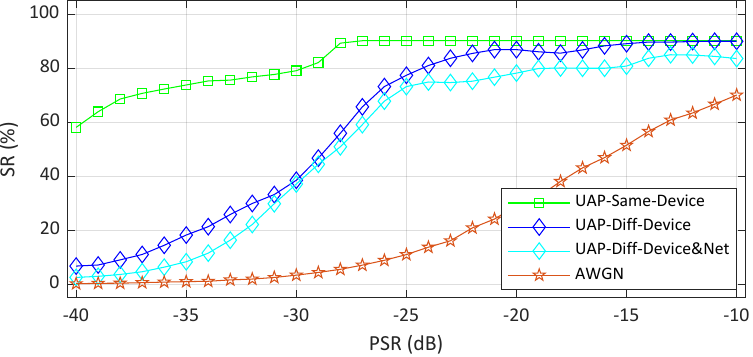}\label{fig:GRUDIFDEC}}
\centering
\caption{SR of UAP when adversary used different devices from the victim systems. (a) CNN1*-based RFFI, (b) LSTM1*-based RFFI, and (c) GRU1*-based RFFI.}
\label{fig:uap_crossdevice}
\end{figure}

Figures~\ref{fig:uap_crossdevice}(a)/(b)/(c) show similar trends among CNN/LSTM/GRU. Without loss of generality, we selected Fig.~\ref{fig:uap_crossdevice}(c) for detailed analysis.

Fig.~\ref{fig:uap_crossdevice}(c) shows the SRs of UAP against GRU1*-based RFFI. 
``UAP-Diff-Device" shows a better performance than ``UAP-Diff-Device\&Net", which indicates that as the adversary gains more knowledge about the victim, SR also increases.
It can be found that over 75\% of the devices are misclassified with ``UAP-Diff-Device\&Net" in the GRU1*-based RFFI system when the PSR is $-25$~dB, and the SR of UAP is significantly higher than the misclassification rate due to AWGN.

When the adversary used different devices from the victim systems, the attack performance did not degrade much from the case when they have the same set of devices (``UAP-Same-Device''). This poses a serious security concern to the RFFI systems as the adversary can build its own testbed for attacks as long as it knows the wireless protocol of the victim system.

\subsection{Real-Time Adversarial Attack}\label{sec:synUAP} 
In practice, in order to inject perturbation into the air, the adversary needs to first detect the transmission and then align to it. The adversary can leverage the repeated preambles in the packet for time synchronization, in the same way as the legitimate receiver does~\cite{shen2021jsac}, because the preamble is publicly known. However, the adversary will not be able to inject perturbation before the synchronization is completed.
The influence on the synchronization of adversarial attacks against RFFI is unclear yet, and it will be discussed in this subsection.

\subsubsection{Setup}
The surrogate dataset and the victim dataset were taken from the same devices (Group 1 devices) on different days. 
The victim models were the models detailed in 
Section~\ref{sec:dlmodel}  (CNN1, LSTM1, GRU1), and the surrogate models are the models in Section~\ref{sec:UAPsetup}  (CNN2, LSTM2, GRU2).

\begin{figure}[!t]
\centering
\includegraphics[width=3.4in]{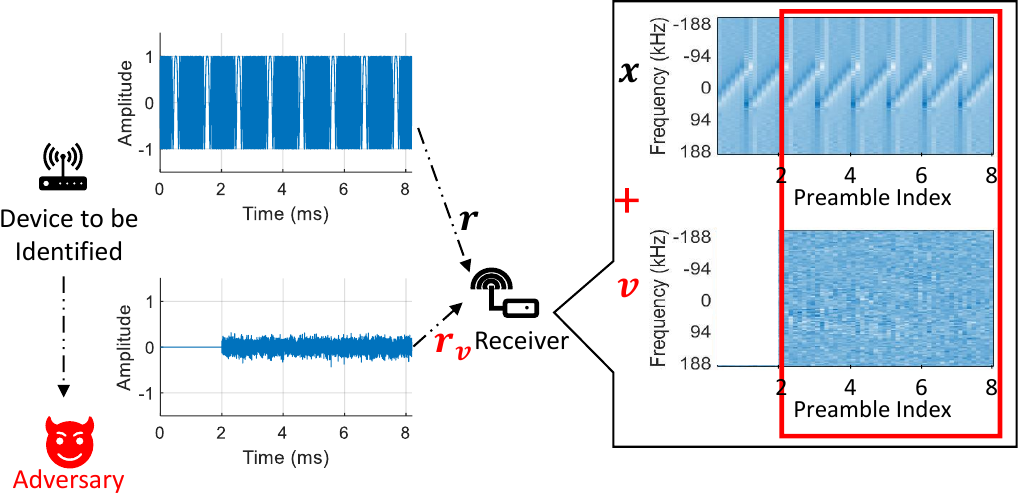}
\caption{Real-time adversarial attacks.}
\label{fig:synattack}
\end{figure}

As shown in Fig.~\ref{fig:synattack}, there are eight repeating preambles in a LoRa packet. When a DUT starts to transmit a LoRa packet, the adversary can perform packet synchronization using the first two LoRa preambles~\cite{shen2021jsac}, in the same way as a legitimate receiver. Once the packet is detected, the adversary will switch to the transmit mode and inject perturbation into the air.

To better meet the practical conditions in wireless communication, we introduced an additional delay  after synchronization ranging from 0 to 1 millisecond. The additional interval is determined based on the different arrival time of transmission paths, “victim-receiver” and “victim-adversary-receiver”, as well as any processing time incurred at the adversary.
The perturbation is superimposed after the synchronization and additional delay, this is termed ``\textbf{Sync-UAP}''.

We also studied three benchmark strategies.
\begin{itemize}
    \item \textbf{Whole UAP:} it is the ideal scenario for the adversary. It refers to the complete UAP added to the victim signals and aligned perfectly.
    \item \textbf{Unsync UAP:} the adversary continuously injects the  UAP signals into the air, without performing any synchronization, i.e., the UAP may not be aligned with the received signal.
    \item  \textbf{AWGN:}
    instead of transmitting UAP, the adversary keeps sending AWGN to the receiver. 
\end{itemize}


\subsubsection{Result}
\begin{figure}[!t]
\centering
\subfloat[]{\includegraphics[width=3.4in]{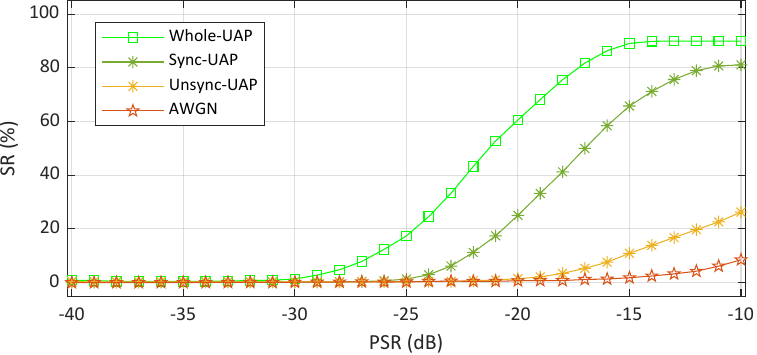}\label{fig:CNNsys}}

\subfloat[]{\includegraphics[width=3.4in]{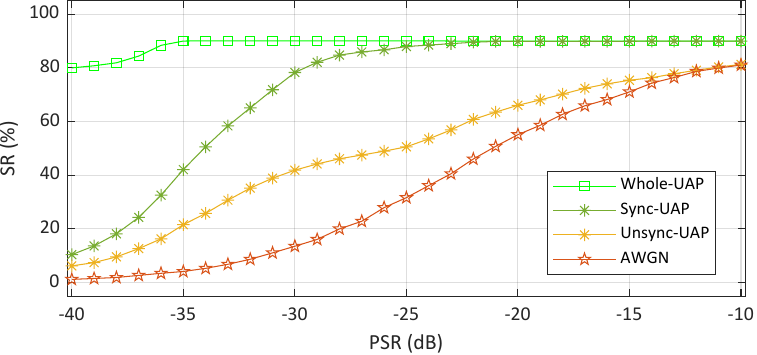}\label{fig:LSTMsys}}\hspace{0.1cm}

\subfloat[]{\includegraphics[width=3.4in]{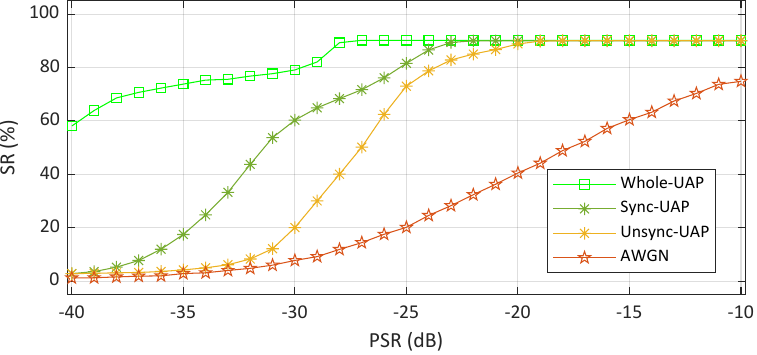}\label{fig:GRUsys}}
\centering
\caption{SR of UAP considering real-time synchronization of perturbation. (a) CNN1-based RFFI. (b) LSTM1-based RFFI. and (c) GRU1-based RFFI.}
\label{fig:bb}
\end{figure}

Fig.~\ref{fig:bb} shows the performance of UAP employing different attack strategies in RFFI. The experiments of ``Sync-UAP'' and ``Unsync-UAP'' were repeated 10 times and the averaged results were provided.
As expected, the ``Whole-UAP'' achieved the best performance.
The degradation between ``Sync-UAP'' and ``Whole-UAP'' can be attributed to the lack of perturbation superimposed on the inputs during the synchronization and propagation.
Although ``Sync-UAP'' does not cover the entire victim signal, it still achieves better performance compared to ``Unsync-UAP'', which indicates the importance of synchronization.
It can be observed that the adversary utilizing ``Sync-UAP'' can approach ``Whole-UAP'' by boosting PSR, which demonstrates that it is effective to launch a UAP attack in real-time.

\subsection{Attack in Practical Settings}\label{sec:ChallengingSce} 
In this section, we will combine the above-studied scenarios to reveal the UAP attacks in a realistic environment.

\subsubsection{Setup}
The setup of the adversary and victim RFFI systems is given below.
\begin{itemize}
	\item \textbf{Adversary} builds a testbed using Group 1 LoRa devices. The surrogate models are CNN2, LSTM2, and GRU1 (described in Section~\ref{sec:dlmodelsetup} and Section~\ref{sec:UAPsetup}). Sync-UAP is employed as it is practical and effective.
	\item \textbf{Victim RFFI systems} involve Group 2 LoRa devices. The DL models are CNN1*, LSTM1*, and GRU1* (mentioned in Section~\ref{sec:cross-devicesetup}).
\end{itemize}

The dataset $\mathcal{D}^{d1}_{G1}$ is used to generate perturbations. These are superimposed onto $\hat{\mathcal{D}}^{\hat{d1}}_{G2}$, $\hat{\mathcal{D}}^{\hat{d2}}_{G2}$, and $\hat{\mathcal{D}}^{\hat{d3}}_{G2}$, which are from Group 2 on January 29$^{\text{th}}$, 30$^{\text{th}}$, and 31$^{\text{st}}$, 2023, respectively.

\subsubsection{Result}
\begin{figure}[!t]
\centering
\subfloat[]{\includegraphics[width=3.4in]{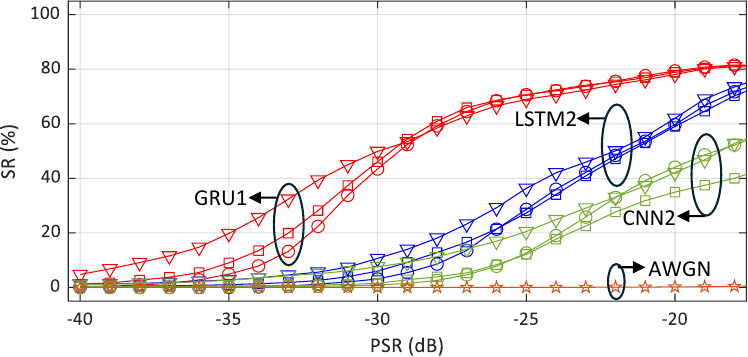}\label{fig:CNNchallenge}}

\subfloat[]{\includegraphics[width=3.4in]{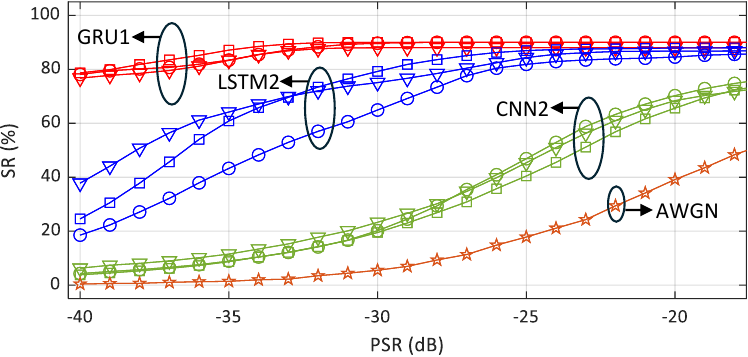}\label{fig:LSTMchallenge}}\hspace{0.1cm}

\subfloat[]{\includegraphics[width=3.4in]{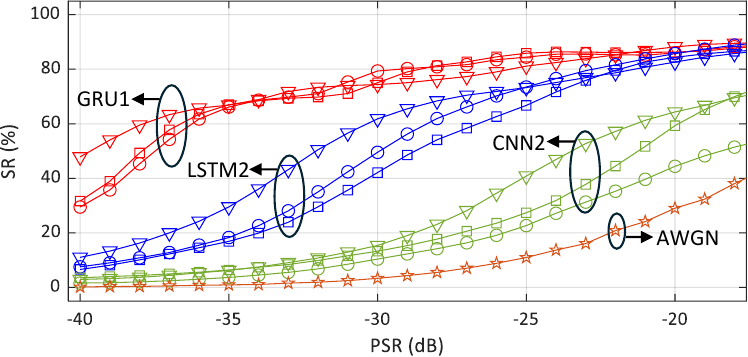}\label{fig:GRUchallenge}}
\centering
\caption{SR of UAP in practical settings. (a) CNN1*-based RFFI. (b) LSTM1*-based RFFI. and (c) GRU1*-based RFFI. 
$\triangledown$ represents day 1, $\square$ represents day 2, $\circ$ represents day 3.} 
\label{fig:challenge}
\end{figure}


Fig.~\ref{fig:challenge}(a) shows the results of the UAPs on CNN1* (Group 2) when the surrogate DL model is CNN2/LSTM2/GRU1 (Group 1).
When $\emph{PSR}=$ -18~dB, superimposing the UAP generated by GRU1 to CNN1* causes SR of 81.7\%, which is 81.3\% higher than the SR achieved with AWGN.
From Fig.~\ref{fig:challenge} we can find that UAP generated by the system  where the surrogate model and devices are different from the victim is still more effective than AWGN, regardless of whether the victim is CNN, LSTM, or GRU. It can also be observed the attack performance is stable over different days.

\section{Conclusion}\label{sec:conclusion}
In this paper, we carried out a comprehensive study of the adversarial attacks on DL-based RFFI and explored the attack performance under practical wireless contexts. A LoRa-RFFI testbed was built and real datasets were collected for experimental evaluation.
Three different adversarial attack methods, i.e., FGSM, PGD, and UAP, were studied. 
These attacks were applied to CNN, LSTM, and GRU-based RFFI systems.
Experimental results show FGSM and PGD in white-box attacks can make packets be classified incorrectly with rates of 87.2\% and 96.2\%, respectively. The iterative method PGD was demonstrated to be more powerful and capable of achieving SR of 98.9\% in a targeted attack.
A more universal attack method, UAP, was conducted to show its effectiveness in grey/black-box attacks and in practical wireless contexts involving unknown DL models, different attack times, inaccessible victim devices, and real-time injection of perturbation. The experimental results depicted a misclassification of 80.5\% of packets in a setting that encompasses the factors mentioned earlier.
In short, the successes of adversarial attacks signify a real threat to the use of RFFI for wireless security and call for prompt efforts of designing efficient countermeasures.

\bibliographystyle{IEEEtran}
\bibliography{IEEEabrv,mybibfile}

\end{document}